\documentclass[a4paper,11pt]{article}
\usepackage[normalem]{ulem}
\usepackage{xcolor}

\newif\ifshowchanges
\showchangesfalse  

\usepackage{xcolor}
\definecolor{myadd}{RGB}{0,100,200}      
\definecolor{mydel}{RGB}{180,30,30}      
\definecolor{myrep}{RGB}{0,150,0}

\ifshowchanges
  \usepackage[normalem]{ulem} % 
  \newcommand{\add}[1]{\textcolor{myadd}{#1}}
  \newcommand{\del}[1]{\textcolor{mydel}{\sout{#1}}}
  \newcommand{\rep}[2]{\textcolor{myrep}{#1}\textcolor{mydel}{\sout{#2}}}
\else
  \newcommand{\add}[1]{#1}   
  \newcommand{\del}[1]{}     
  \newcommand{\rep}[2]{#1}   
\fi

\pdfoutput=1 % if your are submitting a pdflatex (i.e. if you have
\usepackage{jcappub} % for details on the use of the package, please
\usepackage[T1]{fontenc} % if needed
\usepackage{graphicx} % Required for inserting images
\usepackage{amsmath}
\usepackage{bm}
\usepackage{amssymb}
\usepackage{natbib}
\usepackage{booktabs}
\usepackage{subcaption} % 子图支持
\usepackage{float}
\usepackage[utf8]{inputenc}
\usepackage[T1]{fontenc}
\usepackage{siunitx} %专业单位排版
\usepackage{comment}
\usepackage{lineno} 

\title{\boldmath Optimizing Gaussian Process Kernels Using Nested Sampling and ABC Rejection for $H(z)$ Reconstruction}

\author[a,b]{Jia-yan Jiang,}
\author[c,1]{Kang Jiao,\note{Corresponding author.}}
\author[a,b,1]{Tong-Jie Zhang.}

\affiliation[a]{Institute for Frontiers in Astronomy and Astrophysics, Beijing Normal University, Beijing 102206, People’s Republic of China}
\affiliation[b]{School of Physics and Astronomy, Beijing Normal University, Beijing 100875, People’s Republic of China}
\affiliation[c]{Institute for Astrophysics, School of Physics, Zhengzhou University, Zhengzhou 450001, China}

\emailAdd{kangjiao@zzu.edu.cn}
\emailAdd{tjzhang@bnu.edu.cn}

\abstract{Recent cosmological observations have achieved high-precision measurements of the Universe's expansion history, prompting the use of nonparametric methods such as Gaussian processes (GP) regression. We apply GP regression for reconstructing the Hubble parameter using CC data, with improved covariance modeling and latest study in CC data. \add{In addition, we perform a joint analysis combining BAO, SN Ia, and a CMB prior, which constitutes the standard and most constraining framework in cosmology.} By comparing reconstructions in redshift space \( z \) and transformed space \( \log(z+1) \) , we evaluate six kernel functions using nested sampling (NS) and approximate Bayesian computation rejection (ABC rejection) methods and analyze the construction of Hubble constant $H_0$ in different models. Our analysis demonstrates that reconstructions in $\log(z+1)$ space remain physically reasonable, offering a viable alternative to conventional $z$ space approaches, while the introduction of non-diagonal covariance matrices \add{in CC data} leads to degraded reconstruction quality, suggesting that simplified diagonal forms may be preferable for reconstruction. \add{And we find robust evidence for an observable deviation from the $\Lambda$CDM model under the joint constraints of BAO and SNe Ia data with a CMB prior.} These findings underscore the importance of task-specific kernel selection in GP-based cosmological inference. In particular, our findings suggest that careful preliminary screening of kernel functions, based on the physical quantities of interest, is essential for reliable inference in cosmological research using GP.}

\begin{document}
%\linenumbers 
\maketitle
\flushbottom

\section{Introduction}\label{sec:intro}
Gaussian process (GP) has become a powerful and widely adopted tool in modern cosmology, enabling model-independent reconstructions of key cosmological observables. GP avoids prior assumptions about functional forms that may bias cosmological parameter inference \citep{2012JCAP...06..036S,bonilla2021measurements}, offering a more flexible and robust analysis framework. Foundational studies demonstrated the effectiveness of GP in reconstructing the Hubble and deceleration parameters \cite{seikelReconstructionDarkEnergy2012,busti2014evidence,Busti_Clarkson_Seikel_2014}. More recent works reconstructed the dark energy potential in GP \cite{Jesus_2022}, and applied GP to address the Hubble tension \cite{gomez2018h0,mehrabi2020does}, further underscoring their importance in contemporary cosmological analyses.
 
In GP modeling, the kernel function selection embodies fundamental assumptions regarding both the smoothness of the underlying process and its correlation structure, with different kernels corresponding to different hypotheses about cosmological parameters in cosmological applications.
Numerous studies have emphasized the importance of kernel selection. For instance, \add{Sun et al.~\cite{sun2021influence} } introduce theoretical constraints on the hyperparameter space of the radial basis function (RBF) kernel \del{\cite{sun2021influence}} and \add{Zhang et al.~\cite{zhangKernelSelectionGaussian2023}} advance this direction by integrating ABC rejection with NS to identify optimal kernels\del{\cite{zhangKernelSelectionGaussian2023}}. However, these investigations have predominantly operated in redshift $z$ space and often assumed diagonal covariance matrices, thereby neglecting potential correlations among the data and the influence of reconstructing space.

In this work, we develop a comprehensive framework for kernel selection and evaluation in GP cosmology, combining NS and ABC rejection in both diagonal and nondiagonal covariance settings. We test six kernel functions---Matérn 3/2, 5/2, 7/2, 9/2, RBF, and Cauchy, which are carefully selected to represent a continuous spectrum of smoothness assumptions ranging from the rough Cauchy kernel to the infinitely differentiable RBF kernel. This selection enables us to comprehensively explore physically plausible behaviors when reconstructing the Hubble parameter \rep{$H(z)$}{H(z)} from cosmic chronometer measurements \add{and the joint dataset of BAO and SN Ia with a CMB prior}.

A key innovation of our study is the extension of traditional analyses in \(z\) space to include parallel reconstructions in \(\log(z+1)\) space. This transformation facilitates a more effective utilization of low-redshift data and offers a complementary perspective on GP modeling. Our findings demonstrate that reconstructions in \(\log(z+1)\) space are not only viable but also yield qualitatively distinct results, which may have important implications for late-time expansion inference and the ongoing $H_0$ tension debate. These methodological differences could provide valuable insights for future cosmological studies.

Our methodology builds upon the \texttt{GaPP} package, which we enhance with NS and ABC rejection algorithms for robust Bayesian evidence estimation. This hybrid approach reinforces the statistical credibility of our conclusions.

The remainder of this paper is organized as follows. Section~\ref{sec:methods} outlines the model construction, GP implementation and evaluation framework, including details of the ABC rejection algorithm and NS procedure. 
Section~\ref{sec:results} presents the comparative results for different kernels and redshift representations. Section~\ref{sec:discussion}, we discuss the broader implications of our findings. Overall, our results suggest that kernel selection should be guided by the specific physical questions being addressed, as different kernels can lead to markedly different cosmological reconstructions. Importantly, this highlights the need to incorporate kernel choice uncertainty into cosmological parameter error budgets to ensure robust statistical inference.

\section{Methods} \label{sec:methods}

\subsection{Data preparation and processing}
 
Cosmic chronometer (CC) data, derived from the relative age differences of passively evolving galaxies, provide a model-independent approach to measuring the Hubble parameter \( H(z) \). As such, they have become an essential tool in constraining cosmological models \citep{jimenezConstrainingCosmologicalParameters2002, morescoUnveilingUniverseEmerging2022}. This technique offers a unique window into the expansion history of the Universe and the properties of dark energy \citep{2012JCAP...06..036S}. In this study, CC data constitute a fundamental component of our analysis framework.

A critical reassessment of the CC method was recently undertaken \cite{ahlstrom2023use}, who revisited the original approach \cite{simonConstraintsRedshiftDependence2005}. The latter derived eight \( H(z) \) measurements with 10--20\% uncertainties using a sample of 32 galaxies. However, The new work systematically re-evaluated the robustness of this methodology and concluded that achieving the reported precision would require unrealistically low uncertainties in galaxy age measurements--on the order of 1--3\%, far below the accepted threshold of 12\%. This finding underscores the inherent difficulty in accurately determining the differential quantity \( \mathrm{d}t/\mathrm{d}z \), which lies at the core of the CC method.

To investigate these limitations, the authors employed both Monte Carlo simulations and GP regression. Their analysis indicated that the original uncertainty estimates were likely overly optimistic and that achieving robust constraints would necessitate a substantially larger galaxy sample (70--280 galaxies). Furthermore, they highlighted that systematic uncertainties arising from the stellar population synthesis models used to estimate galaxy ages introduce additional model dependence. Consequently, the original eight measurements were distilled into two higher-precision estimates deemed more reliable for cosmological inference, which we adopt in this study.

In the CC data, the Hubble parameter $69\pm12$ $ \add{\mathrm{km\,s^{-1}\,Mpc^{-1}}}$ corresponding to a redshift of $0.09$ is referred to as the Hubble constant $H_0$ in the original literature \cite{jimenezConstraintsEquationState2003}. We reformulate their result to express the redshift-dependent Hubble parameter \( H(z) \) by incorporating the standard $\Lambda$CDM cosmological evolution model:
\begin{equation}
H(z) = H_0 \sqrt{\Omega_m(1+z)^3 + \Omega_\Lambda},
\label{eq:hubble_parameter}
\end{equation}
using this relation, we obtain a refined value of the Hubble parameter,
\begin{equation}
    H(z=0.09) = 70.70 \pm 0.09\,\mathrm{km\,s^{-1}\,Mpc^{-1}},
\end{equation}
which is included in our final dataset, as shown in Table~\ref{tab:ccdata}.

\begin{table}[t]
\centering

\begin{tabular}{lllll}
\toprule
$z$ & $H(z)$ & $\sigma_{H(z)}$ & Method  & Reference \\
\midrule
0.07 & 69 & 19.6 & F & \cite{zhang2014four} \\
0.09 & 70.7 & 12.3 & F & \cite{jimenezConstraintsEquationState2003}$\ast$ \\
0.12 & 68.6 & 26.2 & F & \cite{zhang2014four} \\
0.179 & 75 & 4 & D & \cite{moresco2012improved} \\
0.199 & 75 & 5 & D & \cite{moresco2012improved} \\
0.2 & 72.9 & 29.6 & F & \cite{zhang2014four} \\
0.28 & 88.8 & 36.6 & F & \cite{zhang2014four} \\
0.352 & 83 & 14 & D & \cite{moresco2012improved} \\
0.377 & 97.9 & 22.1 & F & \cite{ahlstrom2023use}$\star$ \\
0.38 & 83 & 13.5 & D & \cite{moresco20166} \\
0.4004 & 77 & 10.2 & D & \cite{moresco20166} \\
0.425 & 87.1 & 11.2 & D & \cite{moresco20166} \\
0.445 & 92.8 & 12.9 & D & \cite{moresco20166} \\
0.47 & 89 & 49.6 & F & \cite{ratsimbazafy2017age} \\
0.4783 & 80.9 & 9 & D & \cite{moresco20166} \\
0.48 & 97 & 62 & F & \cite{stern2010cosmic} \\
0.593 & 104 & 13 & D & \cite{moresco2012improved} \\
0.68 & 92 & 8 & D & \cite{moresco2012improved} \\
0.781 & 105 & 12 & D & \cite{moresco2012improved} \\
0.8 & 113.1 & 25.22 & F & \cite{jiao2023new} \\
0.8754 & 125 & 17 & D & \cite{moresco2012improved} \\
0.88 & 90 & 40 & F & \cite{stern2010cosmic} \\
1.037 & 154 & 20 & D & \cite{moresco2012improved} \\
1.26 & 135 & 65 & F & \cite{tomasetti2023new} \\
1.363 & 160 & 33.6 & D & \cite{moresco2015raising} \\
1.364 & 116.6 & 15.29 & F & \cite{ahlstrom2023use}$\star$ \\
1.965 & 186.5 & 50.4 & D & \cite{moresco2015raising} \\
\bottomrule
\end{tabular}
\caption{Cosmic chronometer $H(z)$ measurements. $\star$: original eight data points replaced by two improved estimates. $\ast$: value converted from $H_0$ using the standard redshift evolution model. The "Method" column indicates the technique used to determine the differential age \( \mathrm{d}t \), where "F" denotes the full-spectrum fitting method and "D" denotes the D4000 index method.}
\label{tab:ccdata}
\end{table}
%%这里开始
% \add{In our analysis, we first conduct a detailed and dedicated investigation using cosmic chronometer (CC) data, which serves as the foundational dataset of this work. Building upon this basis, we then incorporat baryon acoustic oscillation (BAO) and type Ia supernova (SNIa) data for broader constraints. Finally, we perform a comprehensive joint analysis combining BAO, SNIa, and a prior from Cosmic Microwave Background (CMB) observations. The BAO measurements are based on the effective redshift $z_{eff}$ and the corresponding values of $D_{H}/r_d$ from TABLE IV of the DESI BAO DR2 release \cite{karim2025desi}. The SNIa dataset is taken from Scolnic et al\cite{scolnic2018complete}, specifically utilizing the 40 binned data points together with their full covariance matrix. The CMB prior is incorporated to constrain the sound horizon scale at radiation drag $r_d$ and is typically derived from the Planck satellite's full-mission data \cite{aghanim2020planck}.}

\add{To strengthen the GP reconstruction beyond CC data, we incorporate three complementary cosmological probes: baryon acoustic oscillations (BAO), type Ia supernovae (SN Ia), and a prior from the cosmic microwave background (CMB). Each of these datasets plays a distinctive role. }

\add{BAO provides a geometric standard ruler for probing the expansion history of the Universe\cite{eisenstein2005detection}, linking the observed clustering of galaxies to the comoving sound horizon at radiation drag $r_d$. For our reconstruction, we adopt the radial BAO measurements in the form of $D_H(z)/r_d = c/[H(z) r_d]$ from TABLE IV of the DESI BAO DR2 release \cite{karim2025desi}.}

\add{SN Ia, as standardizable candles, probe the luminosity distance $D_L(z)$ and thus provide a powerful tracer of the late-time expansion history. 
The discovery of cosmic acceleration was originally based on SN Ia observations \cite{riess1998observational,perlmutter1999measurements}, establishing their pivotal role in modern cosmology. In this work, we use the distance modulus $\mu(z)$ from the Pantheon compilation \cite{scolnic2018complete}, specifically adopting the 40 binned data points together with their full covariance matrix, which allows us to account for correlated uncertainties in the GP reconstruction.}

\add{The CMB prior, derived from Planck full-mission data \cite{aghanim2020planck}, provides the absolute scale of the sound horizon at the baryon drag epoch, $ r_d $. This early-Universe constraint anchors the geometric measurements from BAO and SNe, ensuring consistency with the physics of the early Universe at $ z \sim 1100 $\cite{hu2002cosmic}.}

\add{By combining BAO, SNe, and CMB information within the GP framework, we achieve a balanced and robust non-parametric reconstruction of the Hubble constant, mitigating the limitations of individual datasets while exploiting their complementary sensitivities.}

%这里结束

\subsection{Model construction}
% \add{\subsubsection{CC Data Analysis Framework}}
\add{In CC data,} we reconstruct the Hubble parameter \( H(z) \) in both the native redshift space \( z \) and the logarithmic redshift space \( \log(z+1) \). Performing the reconstruction in \( \log(z+1) \) offers several advantages, particularly when using CC dataset characterized by dense sampling at low redshift (\( z < 0.5 \)) and sparse coverage at higher redshifts (\( z > 1 \)). The transformation \( \log(z+1) \) expands the low-redshift range while compressing the high-redshift regime, thereby alleviating the non-uniformity in data distribution. This transformation reduces the disproportionate influence of densely clustered low-redshift data (with $16$ data points at $z < 0.5$) and enhances the relative contribution of high-redshift points (only $6$ data points at $z > 1$), resulting in a more balanced and robust reconstruction across the full redshift range. Furthermore, the point \( \log(z+1) = 0 \) naturally corresponds to the Hubble constant \( H_0 \), ensuring consistency with standard reconstructions in the \( z \) domain. The logarithmic transformation provides particular advantages for constraining $H_0$, as it naturally weights the low-redshift regime where the Hubble constant is most sensitively probed.

Observational covariances are explicitly incorporated by modeling correlations among data points \cite{moresco2020setting,moresco_cccovariance}. This methodology enabled the derivation of a full 15-point covariance matrix for the D4000-based measurements summarized in Table~\ref{tab:ccdata} in method 'D'. The covariance matrix was obtained from the public repository, and we assume that data points from other methods (e.g., F) and between different method categories (e.g., F vs D) are uncorrelated.

In our analysis, we adopt the full 15-point covariance matrix to more accurately represent the statistical correlations among the data points. This matrix is integrated into our GP framework to improve the fidelity of the covariance structure, thereby enabling more precise and reliable reconstructions of \( H(z) \). For comparison, we also perform reconstructions using a diagonal covariance matrix, assuming uncorrelated uncertainties, in order to evaluate the impact of including full covariance information.

To systematically investigate the effects of redshift representation and covariance structure, we consider four distinct GP models, defined by combinations of the redshift domain (either \( z \) or \( \log(z+1) \)) and the covariance matrix type (full or diagonal). For clarity and to facilitate future reference, we define the corresponding abbreviations in Table~\ref{tab:model}.

\begin{table}[t]
\centering
\begin{tabular}{cll}
\toprule
Redshift Space & Covariance Matrix & Model \\
\midrule
$z$ & Full covariance & Full-z \\
$z$ & Diagonal covariance & Diag-z \\
$\log(z+1)$ & Full covariance & Full-log \\
$\log(z+1)$ & Diagonal covariance & Diag-log \\
\bottomrule
\end{tabular}
\caption{Abbreviations for the four reconstruction models based on different redshift spaces and covariance structures.}
\label{tab:model}
\end{table}

% \subsubsection{\add{BAO and SN Ia Data Analysis Framework}}
\add{In our analysis of both the BAO and SN Ia data, we adopt the original covariance matrices as provided by the respective data releases. Specifically, the BAO data analysis utilize a diagonal covariance matrix, whereas the SN Ia analysis incorporate the full covariance matrix. For each dataset, we perform cosmological reconstruction in both redshift $z$ space and log-transformed redshift $\log(z + 1)$ space. This dual approach is also extended to the final combined analysis.}

\subsection{Gaussian Process}\label{GP}
The Gaussian process (GP) offers a powerful nonparametric framework for reconstructing cosmological functions directly from observational data, extending multivariate Gaussian distributions to function spaces of infinite dimension \citep{rasmussenGaussianProcessesMachine2008}. This probabilistic approach enables flexible modeling of continuous functions without assuming specific parametric forms, making it particularly well-suited for cosmological applications where theoretical models remain uncertain. In cosmology, GP has been extensively applied to
 reconstructing the Hubble parameter $H(z)$, constraining dark energy dynamics, and investigating tensions in cosmic expansion \citep{seikelReconstructionDarkEnergy2012,zhangKernelSelectionGaussian2023,velazquez2024non,biswas2024embedding}.

 The mathematical formulation begins with a dataset $\mathcal{D} = \{(z_i, H_i)\}_{i=1}^n$, comprising redshift measurements $z_i$ and the corresponding Hubble parameter values $H_i = H(z_i) \pm \sigma_i$, where $\sigma_i$ denotes observational uncertainties. A GP is fully characterized by its mean function $\mu(z)$ and covariance kernel $k(z, z')$, expressed as:
\begin{equation}
    f(z) \sim \mathcal{GP}(\mu(z), k(z, z')).
\end{equation}

Following standard practice in cosmological GP analyses \citep{seikelReconstructionDarkEnergy2012}, a zero mean function $\mu(z) = 0$ is typically assumed, allowing the covariance kernel to fully capture the function's behavior. The covariance between observations at redshifts $z_i$ and $z_j$ is determined through kernel functions:
\begin{equation}
    \text{cov}(f(z_i), f(z_j)) = k(z_i, z_j),
\end{equation}
with the covariance matrix for observational set $\bm{Z} = \{z_i\}$ defined by $[K(\bm{Z}, \bm{Z})]_{ij} = k(z_i, z_j)$.

Predictions for new redshifts $\bm{Z}^*$ are derived through GP regression, yielding the posterior mean and covariance.

\begin{align}
    \bar{\bm{f}}^* &= K(\bm{Z}^*, \bm{Z})[K(\bm{Z}, \bm{Z}) + C]^{-1}\bm{H} \label{eq:gp_mean}, \\
    \text{cov}(\bm{f^*}) &= K(\bm{Z}^*, \bm{Z}^*) - K(\bm{Z}^*, \bm{Z})[K(\bm{Z}, \bm{Z}) + C]^{-1}K(\bm{Z}, \bm{Z}^*),
\end{align}
\noindent where $C = \text{diag}(\sigma_i^2)$ is the noise covariance matrix and $\bm{H}$ is the vector of observed Hubble parameter values. \citep{rasmussenGaussianProcessesMachine2008,zhangKernelSelectionGaussian2023}. These expressions yield robust predictions of $H(z)$ at unobserved redshifts, with uncertainties naturally propagated through the GP framework.

The choice of kernel function is critical to reconstruction accuracy. Commonly used kernels in cosmological GP analyses include:

\begin{itemize}
    \item {Radial Basis Function (RBF)}
    \begin{equation}\label{RBF}
    k(z_i,z_j)=\sigma_f^2\exp({-\frac{(z_i-z_j)^2}{2l^2}}),
    \end{equation}
 %known for its smoothness but sometimes overly restrictive when modeling complex cosmological features \citep{seikelReconstructionDarkEnergy2012}.
\end{itemize}

\begin{itemize}
    \item {Cauchy kernel (CHY)}
    \begin{align}\label{CHY}
        k(z_i,z_j)=\sigma_f^2(1+\frac{(z_i-z_j)^2}{2l^2})^{-1},
    \end{align}
    %which offers flexibility for modeling non-smooth functions.
    \item{Matérn($\nu$=3/2) kernel (M32)}
    \begin{align}\label{M32}
        &k(z_i,z_j)=
         \sigma_f^2 \left( 1 + \frac{\sqrt{3}|z_i-z_j|}{\ell} \right) \exp\left( -\frac{\sqrt{3} |z_i-z_j|}{\ell} \right) ,
    \end{align}
    %balancing smoothness and flexibility \citep{biswas2024embedding}.
    \item {Matérn($\nu$=5/2) kernel (M52)}
    \begin{align}\label{M52}
        &k(z_i,z_j)=
        \sigma_f^2\exp({-\sqrt{5}\frac{|z_i-z_j|}{l}})(1+\sqrt{5}\frac{|z_i-z_j|}{l}+5\frac{(z_i-z_j)^2}{3l^2}),
    \end{align}
    %often preferred for its adaptability to cosmological datasets \citep{zhangKernelSelectionGaussian2023}.
    \item {Matérn($\nu$=7/2) kernel (M72)}
    \begin{align}\label{M72}
        &k(z_i,z_j)=
        \sigma_f^2\exp({-\sqrt{7}\frac{|z_i-z_j|}{l}})(1+\sqrt{7}\frac{|z_i-z_j|}{l}+14\frac{(z_i-z_j)^2}{5l^2}+7\sqrt{7}\frac{|z_i-z_j|^3}{15l^3}),
    \end{align}
    %providing higher smoothness for complex reconstruction.
    \item {Matérn($\nu$=9/2) kernel (M92)}
    \begin{align}\label{M92}
        &k(z_i,z_j)=\sigma_f^2\exp({-3\frac{|z_i-z_j|}{l}})
        (1+3\frac{|z_i-z_j|}{l}+27\frac{(z_i-z_j)^2}{7l^2}+18\frac{|z_i-z_j|^3}{7l^3}+27\frac{(z_i-z_j)^4}{35l^4}),
    \end{align}
    %used for highly smooth functions \citep{rasmussenGaussianProcessesMachine2008}.
\end{itemize}
here, $\sigma_f^2$ controls the signal variance, and $l$ represents the characteristic length scale \citep{seikelReconstructionDarkEnergy2012}. Recent studies indicate that Matérn kernels with $\nu = 5/2$ or $7/2$ often outperform the RBF kernel in cosmological reconstructions due to their better adaptability to varied data features \citep{zhangKernelSelectionGaussian2023}.

The hyperparameters $\theta = ( \sigma_f,l)$ are optimized by maximizing the \rep{evidence}{log marginal likelihood}:
\begin{equation}
\begin{aligned}
    \log p(\bm{H}|\bm{Z},\theta) &= -\frac{1}{2}\bm{H}^\top[K(\bm{Z},\bm{Z}) + C]^{-1}\bm{H} \\
    &\quad -\frac{1}{2}\log|K(\bm{Z},\bm{Z}) + C| - \frac{n}{2}\log(2\pi),
\end{aligned}
\label{eq:log_marginal_likelihood}
\end{equation}

\noindent which balances model fit and complexity according to the Bayesian Occam's razor principle \citep{rasmussenGaussianProcessesMachine2008}. This optimization approach has been successfully applied to cosmic chronometer data \citep{zhangKernelSelectionGaussian2023}, dark energy equation of state reconstruction \citep{seikel2013optimising}, and analyses of the Hubble tension \citep{gomez2018h0}.

The nonparametric nature of GP is especially advantageous in cosmology, where theoretical uncertainties remain significant. By avoiding strong assumptions about functional forms, GP enables data-driven exploration of the expansion history of the Universe while naturally propagating observational uncertainties \citep{seikelReconstructionDarkEnergy2012}. 

Recent methodological advances, including ABC rejection, NS, and kernel combination techniques, have further improved model robustness and facilitated systematic kernel selection \citep{zhangKernelSelectionGaussian2023}, thus enhancing the reliability of cosmological inferences.

\subsection{Nested Sampling(NS)}\label{NS}

Nested sampling (NS), originally proposed by \cite{skillingNestedSamplingGeneral2006a}, is a powerful Monte Carlo technique designed to compute Bayesian evidence (also known as the marginal likelihood) efficiently. For a model $ M $ with parameters $ \theta $ and observed data $ D $, the Bayesian evidence is given by:
\begin{equation}
    P(D \mid M) = \mathcal{Z} = \int P(D \mid \theta, M) P(\theta \mid M) \, d\theta.
\end{equation}

In Bayesian inference, the evidence $ P(D \mid M) $ quantifies the probability of the observed data under model $M$, marginalized over the entire parameter space. This quantity plays a central role in model selection, as it enables computation of the Bayes factor \citep{kass1995bayes,jeffreys1998theory}, which measures the relative support of two competing models. For two models $ M_1 $ and $ M_2 $, the Bayes factor is defined as:
\begin{equation}
    \frac{P(M_1 \mid D)}{P(M_2 \mid D)} = \frac{P(D \mid M_1) P(M_1)}{P(D \mid M_2) P(M_2)} = \frac{\mathcal{Z}_1}{\mathcal{Z}_2} \frac{\pi_1}{\pi_2},
\end{equation}
where $ \pi_i $ denotes the prior probability of model $ M_i $.
Assuming equal prior probabilities ($ \pi_1 = \pi_2 $), the Bayes factor reduces to the ratio of the evidences $ \mathcal{Z}_1 /\ \mathcal{Z}_2 $. Thus, NS offers a practical and accurate way to evaluate such model comparisons directly.

The core idea of nested sampling is to transform the multidimensional evidence integral into a one-dimensional integral over the prior volume $X$ \citep{skillingNestedSamplingGeneral2006a}:
\begin{equation}
    \mathcal{Z} = \int_{0}^{1} \mathcal{L}(X) \, dX,
\end{equation}
where $ \mathcal{L}(X) $ is the likelihood corresponding to the remaining prior volume  $ X $. The prior volume itself is defined through:
\begin{equation}
    X(\mathcal{L}) = \int_{\mathcal{L}(\theta) > \mathcal{L}} P(\theta \mid M) \, d\theta.
\end{equation}

The NS algorithm proceeds iteratively: at each step, the point with the lowest likelihood $ \mathcal{L}_i $ is removed and replaced with a new point sampled from the prior under the constraint $\mathcal{L}(\theta) > \mathcal{L}_i$. The prior volume is updated accordingly. The evidence is then approximated as a weighted sum:
\begin{equation}
    \mathcal{Z} \approx \sum_{i} \mathcal{L}_i \Delta X_i,
\end{equation}
where $ \Delta X_i $ denotes the contraction in prior volume at iteration $i$. This process continues until the remaining prior volume contributes negligibly to the total evidence.

As demonstrated in \cite{zhangKernelSelectionGaussian2023}, the \rep{evidence}{log marginal likelihood (LML)} of GP models can be directly used as the likelihood function within NS. This enables simultaneous hyperparameter optimization and model comparison across different kernel choices. In our analysis, we employ the \texttt{Dynesty} package \citep{speagleDynestyDynamicNested2020}, which implements NS using dynamic allocation, multi-ellipsoidal decomposition, and random-walk sampling, facilitating efficient exploration of parameter space.

Uniform priors are adopted for all kernel hyperparameters to ensure fair comparisons, with $\sigma_f \in [50, 500]$ and $l \in [0.1, 10]$. These ranges are carefully selected to fully encompass the optimal hyperparameters of all considered kernel functions while maintaining physically plausible scales. The number of live points is chosen to be sufficiently large, which we find adequate for our two-dimensional optimization problem. From the resulting weighted posterior samples, we compute the posterior mean and standard deviation of $H(z=0)$ (or equivalently $H(\log(z+1))=0$).

The Bayesian evidence calculated for each kernel serves as the criterion for model selection. The relative support for model $M_1$ over $M_2$ is quantified by the log Bayes factor:
\begin{equation}
    \ln B_{12} = \ln \mathcal{Z}_1 - \ln \mathcal{Z}_2,
\end{equation}
where $\ln \mathcal{Z}_i$ denotes the natural logarithm of the evidence for model $M_i$, as provided by \texttt{Dynesty}.

We assess the strength of evidence based on the commonly used Jeffreys’ scale \citep{jeffreys1998theory,sarro2012astrostatistics}. Table~\ref{tab:evidence_strength} summarizes the interpretation of $\ln B_{12}$ in terms of odds, \rep{evidence}{posterior probabilities}, and qualitative strength of evidence, assuming equal prior model probabilities $P(M_1) = P(M_2)$.

\begin{table}[t]
\centering

\begin{tabular}{|l|l|l|l|}
\hline
$ \ln B_{12} $ & Odds & Probability &Strength of Evidence \\
\hline
$ < 1.0 $ & $ \lesssim 3:1 $ & $\lesssim 0.750$ & Inconclusive \\
1.0 & $ \sim 3:1 $ & $0.750$ &Weak evidence \\
2.5 & $ \sim 12:1 $ & $0.923$ &Moderate evidence \\
5.0 & $ \sim 150:1 $ & $0.993$ &Strong evidence \\
\hline
\end{tabular}
\caption{Interpretation of Bayes factor strength for model comparison based on the logarithmic Bayes factor $\ln B_{12} = \ln ( \mathcal{Z}_1 / \mathcal{Z}_2 )$, following Jeffreys' scale.}
\label{tab:evidence_strength}
\end{table}

\subsection{Approximate Bayesian Computation(ABC) Rejection}\label{ABC}

The Approximate Bayesian Computation (ABC) rejection method provides a likelihood-free approach to approximating the posterior distribution $ P(\theta|y) $ by drawing samples from the prior and accepting only those that yield simulated data sufficiently close to the observed data. According to Bayes' theorem \citep{stone2013bayes}, the posterior is given by:

\begin{equation}
P(\theta|y) = \frac{P(\theta)P(y|\theta)}{\int P(y|\theta)P(\theta)d\theta},
\label{bayes_eq}
\end{equation}
where $ P(\theta) $ is the prior distribution, $ P(y|\theta) $ is the likelihood function, and the denominator represents the model evidence. For convenience, we typically set $\int P(y|\theta)P(\theta)d\theta =1$ when working with unnormalized posteriors. In our context, $ y $ refers to the observed Hubble parameter data $ H(z) $, while $ \theta $ denotes the kernel type $ M $ and its associated hyperparameters $ (\sigma_f, l) $.

Due to the intractability of the likelihood function $ P(y|\theta) $ in GP models, we employ the ABC rejection algorithm to approximate the posterior distribution. The approximation is given by:

\begin{equation}
P(\theta|y) \approx P(\theta)P(d(y, \bar{f}) \leq \epsilon),
\end{equation}
where $ \bar{f} $ is the GP-reconstructed function obtained from  Eq.~\eqref{eq:gp_mean} with the given hyperparameters $ \sigma_f $ and $l$, $ d(y, \bar{f}) $ is a distance metric that quantifying the discrepancy between the observed data $y$ and the reconstructed function $ \bar{f} $, and $ \epsilon $ is a predefined tolerance threshold.

This yields an approximate posterior distribution over the kernel type $M$ and its associated hyperparameters:

\begin{equation}
P(M, \theta|y) = P(M, \theta)P(d(y, \bar{f}) \leq \epsilon).
\end{equation}

The choice of distance function $ d(y,\bar{f})$ plays a critical role in ABC methods \citep{abdessalem2017automatic,bernardo2021towards}. While the \rep{evidence}{log marginal likelihood} is often used in GP-based inference, it has already been adopted as the primary criterion in our NS framework. To ensure methodological complementarity, we instead adopt the chi-squared statistic $\chi^2$ as the distance function $ d(y, \bar{f}^*) $ , a metric widely used in cosmological data analysis \citep{conley2010supernova,10.1214/16-BA1002,zhangKernelSelectionGaussian2023,bernardo2021towards}. Despite its simplicity and interpretability, $\chi^2$ remains a pragmatic and widely adopted choice for model evaluation. Its computational efficiency and direct connection to likelihood-based inference make it especially suitable for nonparametric frameworks with limited sample sizes. The chi-squared distance is defined as:
\begin{equation}
    \chi^2 = (y - \bar{f}^*)^T C ^{-1} (y - \bar{f}^*),
    \label{x^2}
\end{equation}
where $\bar{f}^*$ is the GP prediction, and $C$ is the covariance matrix of the observations. This measure effectively captures the deviation between the observed data and the model predictions, accounting for data uncertainties.

For each candidate kernel, we perform ABC rejection sampling with a total of $ 10^6 $ iterations. To ensure numerical stability and representativeness, we retain only the last $10^3$ accepted samples from every batch $10^4$ iterations to compute the acceptance rate and estimate the posterior distribution. The sampling stability is ensured by monitoring batch-wise acceptance rates, where we require the relative standard deviation (RSD) of acceptance rates across all $10^3$ batches (each with $10^4$ iterations) to be less than 0.05\%, ensuring that both the posterior shape and key summary statistics (e.g., mean $H(z=0)$ or $H(\log(z+1)=0)$ remain stable. Finally, the acceptance rates across all kernel functions are normalized to yield the relative posterior probabilities for each kernel. \add{Finally, the acceptance rates across all kernel functions are normalized to yield the relative posterior probabilities for each kernel, and we define the normalized posterior probability as}
\add{\begin{equation}
    P(M_i \mid y) = \frac{N_i}{\sum_j N_j},
    \label{eq:nn}
\end{equation}
where $N_i$ denotes the number of accepted samples corresponding to kernel $M_i$.
This normalized posterior probability provides a relative measure of model support across kernels within the ABC rejection framework. It should be emphasized that, unlike the evidence obtained from nested sampling, Eq.~\eqref{eq:nn} does not represent an absolute Bayesian evidence but rather an empirical approximation derived from acceptance frequencies.}

To facilitate model comparison, we again employ the Bayes factor, which can be approximated in the ABC rejection framework as:
\begin{equation}
B_{12} = \frac{p(y|\theta_1)}{p(y|\theta_2)} = \frac{p(\theta_1|y)}{p(\theta_1)}\bigg/\frac{p(\theta_2|y)}{p(\theta_2)} \approx \frac{p(d(y,\bar{f}_1) \le \epsilon)}{p(d(y,\bar{f}_2) \le \epsilon)}.
\end{equation}

The interpretation of Bayes factors in this context follows the conventional Jeffreys' scale \citep{jeffreys1998theory}, as summarized in Table~\ref{tab:bayes_grades}.

\begin{table}[t]
\centering

\begin{tabular}{|c|c|}
\hline
 Range of \( B_{12} \) & Interpretation \\
\hline
 $ < 10^{-1} $ & negative \\
 $ 10^{-1}$  to $10^{\frac{1}{2}} $ & barely worth mention \\
 $ 10^{\frac{1}{2}}$ to $10^{1 }$ &  substantial \\
 $ 10^{1} $   to $10^{\frac{3}{2}} $ &  strong \\
 $ 10^{\frac{3}{2}}$ to $10^{2} $ & very strong \\
 $ > 10^{2}  $ &  decisive \\
\hline
\end{tabular}
\caption{Evidence strength based on Bayes factor \( B_{12} \) in ABC rejection.}
\label{tab:bayes_grades}
\end{table}

\section{Results Analysis} \label{sec:results}

\subsection{Results from NS Analysis \add{with CC data}} \label{sec:nested_results}
\begin{figure}[H]
    \centering
    \includegraphics[width=0.6\linewidth]{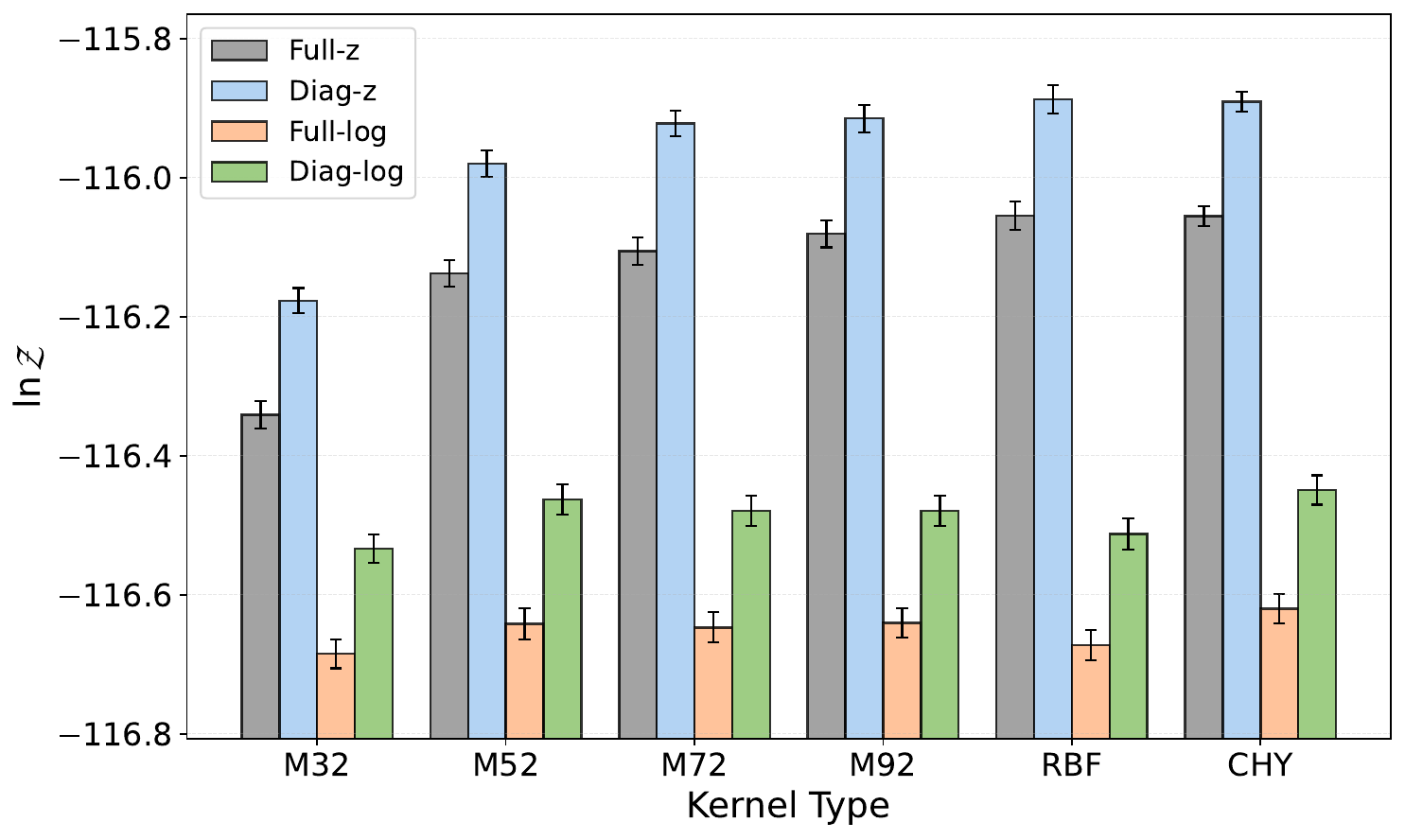}
    \caption{\rep{Evidence comparison across kernel functions for the four reconstruction models defined in Table \ref{tab:model}, obtained from nested sampling of the CC data. Error bars indicate the 1$\sigma$ uncertainties in the nested sampling estimates.}{Log marginal likelihood comparison across kernel functions in four modeling framework. The horizontal axis indicates different kernel types, while the vertical axis shows the corresponding LML values.}}
    \label{fig:ns}
\end{figure}

We \rep{conduct}{conducte} a Bayesian model comparison on the CC data using GP regression, implemented via the \texttt{Dynesty} NS package. Table~\ref{tab:NS} presents the \rep{evidence}{LML} values $\ln\mathcal{Z}$ for four distinct modeling frameworks in Table~\ref{tab:model}, each evaluated with six different kernel functions. Notably, the differences in \rep{evidence}{LML} values across kernels are remarkably small ($\Delta \ln\mathcal{Z} < 1$), suggesting that the current CC dataset lacks sufficient constraining power to strongly discriminate between kernel choices. Figure~\ref{fig:ns} provides a visual summary of the data presented in Table~\ref{tab:NS}.
 
As shown in Figure~\ref{fig:ns}, the log-evidence ($\ln \mathcal{Z}$) computed in $\log(z+1)$ space is only marginally lower (about 0.4) compared to the results of the $z$ space. This close agreement demonstrates the validity of $\log(z+1)$ space reconstruction, providing a solid basis for our subsequent investigations. A key finding from Figure~\ref{fig:ns} is that optimal kernel selection exhibits a stronger dependence on the input space transformation than on the structure of the covariance matrix. Specifically, the RBF kernel achieves optimal performance in $z$ space, while the CHY kernel yields superior evidence in $\log(z+1)$ space.

\begin{table}[t]
\centering

\begin{tabular}{lllll}
\toprule
Model & Kernels & $ln \mathcal{Z}$ \\
\midrule
Full-z& M32&$-116.341 \pm 0.019$\\
&M52& $-116.138 \pm 0.019 $\\
&M72& $-116.105 \pm 0.020 $\\
&M92& $-116.081 \pm 0.019 $\\
&RBF& $-116.055 \pm 0.020 $\\
&CHY& $-116.055 \pm 0.014 $\\
\midrule
Diag-z &M32 & $-116.177 \pm 0.018$\\
&M52& $-115.980 \pm 0.019$\\
&M72& $-115.922 \pm 0.019$\\
&M92& $-115.915 \pm 0.020$\\
&RBF& $-115.887\pm 0.020$\\
&CHY& $-115.891\pm 0.014$\\
\midrule
Full-log &M32&$-116.685 \pm 0.021$\\
&M52& $-116.642 \pm 0.022$\\
&M72& $-116.647 \pm 0.021$\\
&M92& $-116.641 \pm 0.021$\\
&RBF& $-116.672\pm 0.021 $\\
&CHY& $-116.620 \pm 0.021$\\
\midrule
Diag-log&M32&$-116.534\pm 0.021$\\
&M52& $-116.463 \pm 0.022$\\
&M72& $-116.479\pm 0.022$\\
&M92& $-116.479\pm 0.021$\\
&RBF& $-116.512\pm 0.022$\\
&CHY& $-116.449\pm 0.021$\\

\bottomrule

\end{tabular}
\caption{The Evidence $\ln \mathcal{Z}$ obtained for four models using various kernels \add{in NS}.}
\label{tab:NS}
\end{table}

Furthermore, we consistently observe that diagonal covariance models outperform their full covariance counterparts across all configurations. This finding naturally raises questions about the underlying reasons for the superiority of diagonal covariance models, which assume negligible off-diagonal correlations. This apparent advantage warrants careful interpretation. Given the limited dataset size of only 27 measurements—with just 15 D4000-based points for which pairwise covariances are modeled—the full covariance matrix is inherently sparse, with most off-diagonal elements set to zero. Notably, non-zero off-diagonal terms appear only within the D-method subset, with values typically smaller than $40$, while all cross-method correlations (e.g., between F and D) are assumed to be zero. As such, the full covariance structure effectively reduces to a block-diagonal form, dominated by a small D-method block and surrounded by diagonal entries elsewhere. This configuration likely limits the gain from accounting for off-diagonal terms. To mitigate such instabilities, we employ Cholesky decomposition for matrix operations, which ensures numerical stability during GP regression even in the presence of near-singular or ill-conditioned covariance structures. Consequently, diagonal models may outperform simply because they avoid amplifying uncertainties from weakly constrained covariance entries. Further investigation using larger datasets with denser and more robust covariance estimates—especially those incorporating cross-method correlations—could help determine whether this observed advantage generalizes beyond our current setting.
\begin{figure}[H]
    \centering
    \noindent % 取消缩进（可选）
    % 第一行
    \begin{subfigure}[b]{0.39\textwidth} % 减小宽度
        \includegraphics[width=\linewidth]{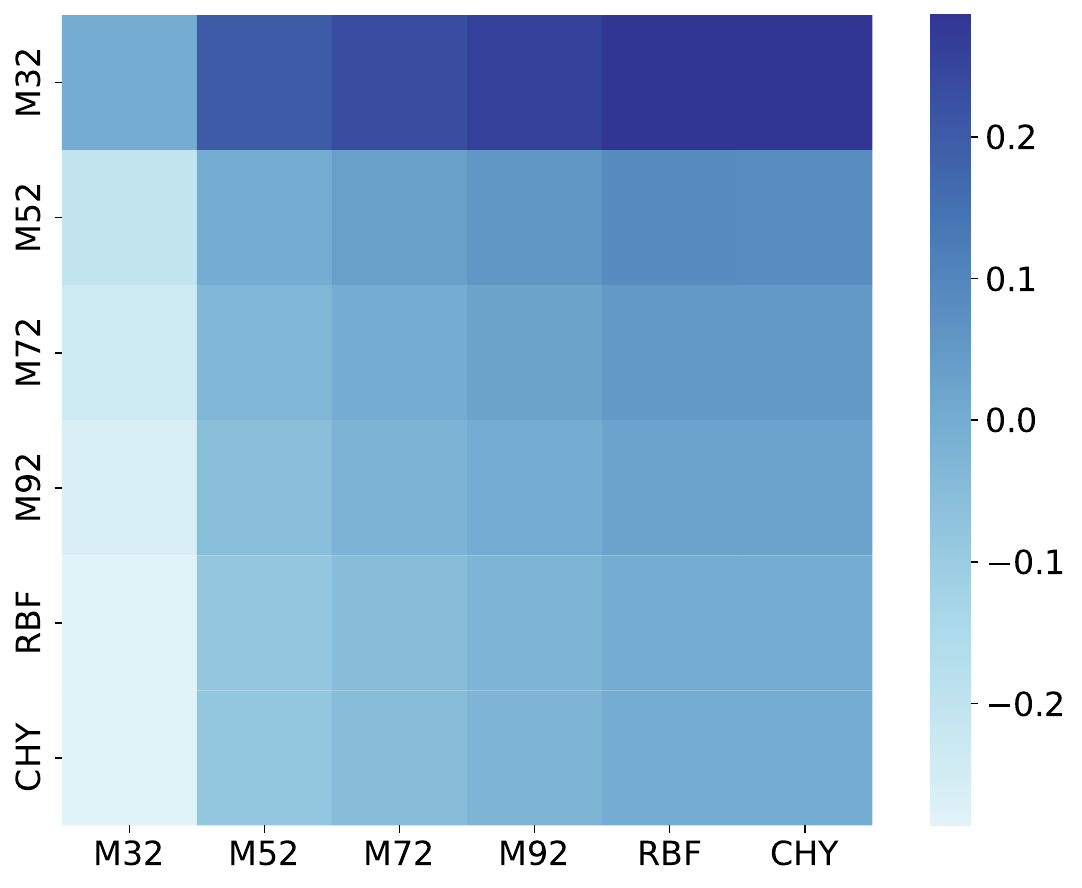}
        \caption{Full-z}
        \label{fig-ns-sub1}
    \end{subfigure}
    \hfill
    \begin{subfigure}[b]{0.39\textwidth}
        \includegraphics[width=\linewidth]{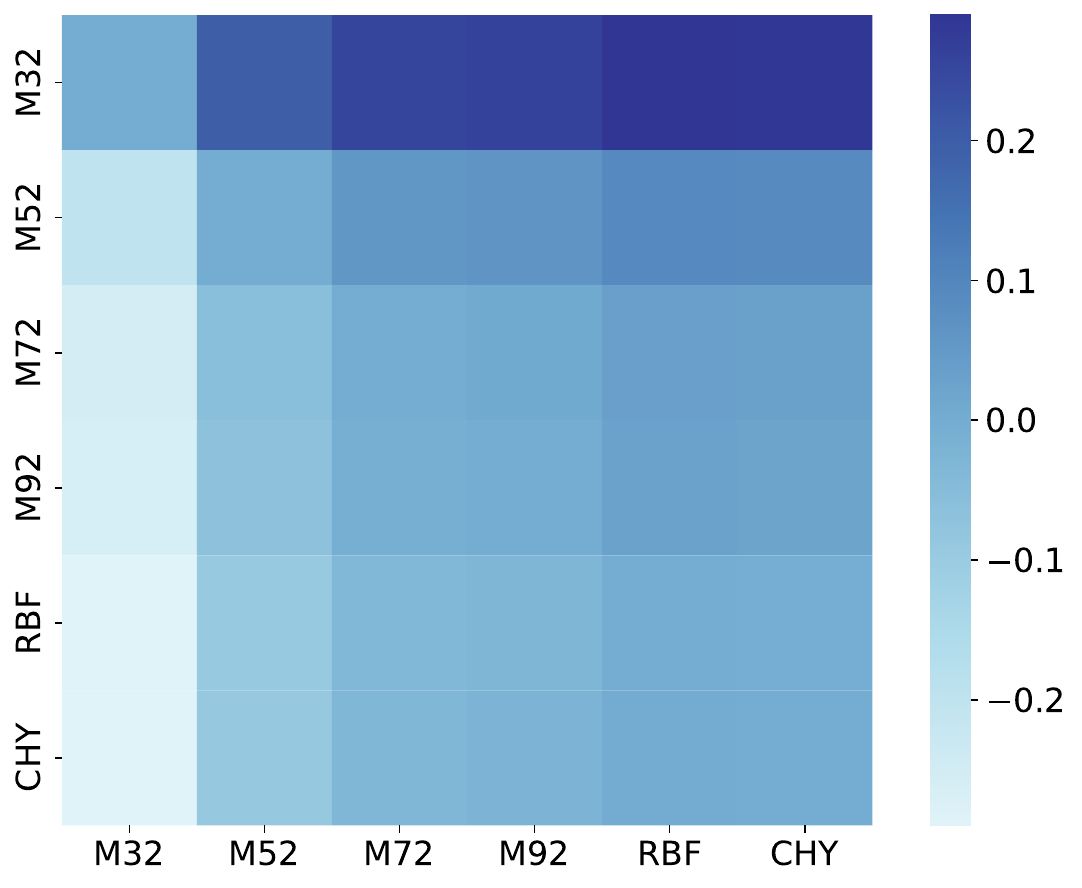}
        \caption{Diag-z}
        \label{fig-ns-sub2}
    \end{subfigure}
    
    % 第二行
    \vspace{0.5em}
    \begin{subfigure}[b]{0.39\textwidth}
        \includegraphics[width=\linewidth]{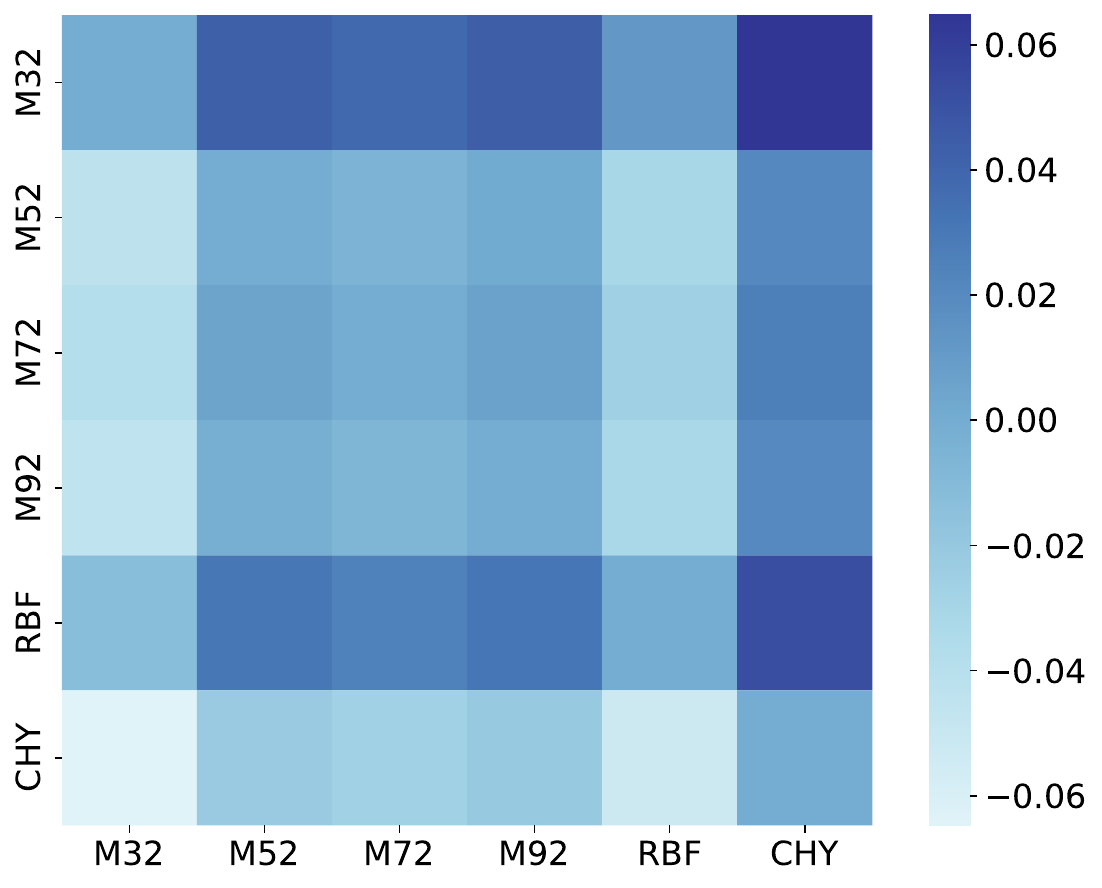}
        \caption{Full-log}
        \label{fig-ns-sub3}
    \end{subfigure}
    \hfill
    \begin{subfigure}[b]{0.39\textwidth}
        \includegraphics[width=\linewidth]{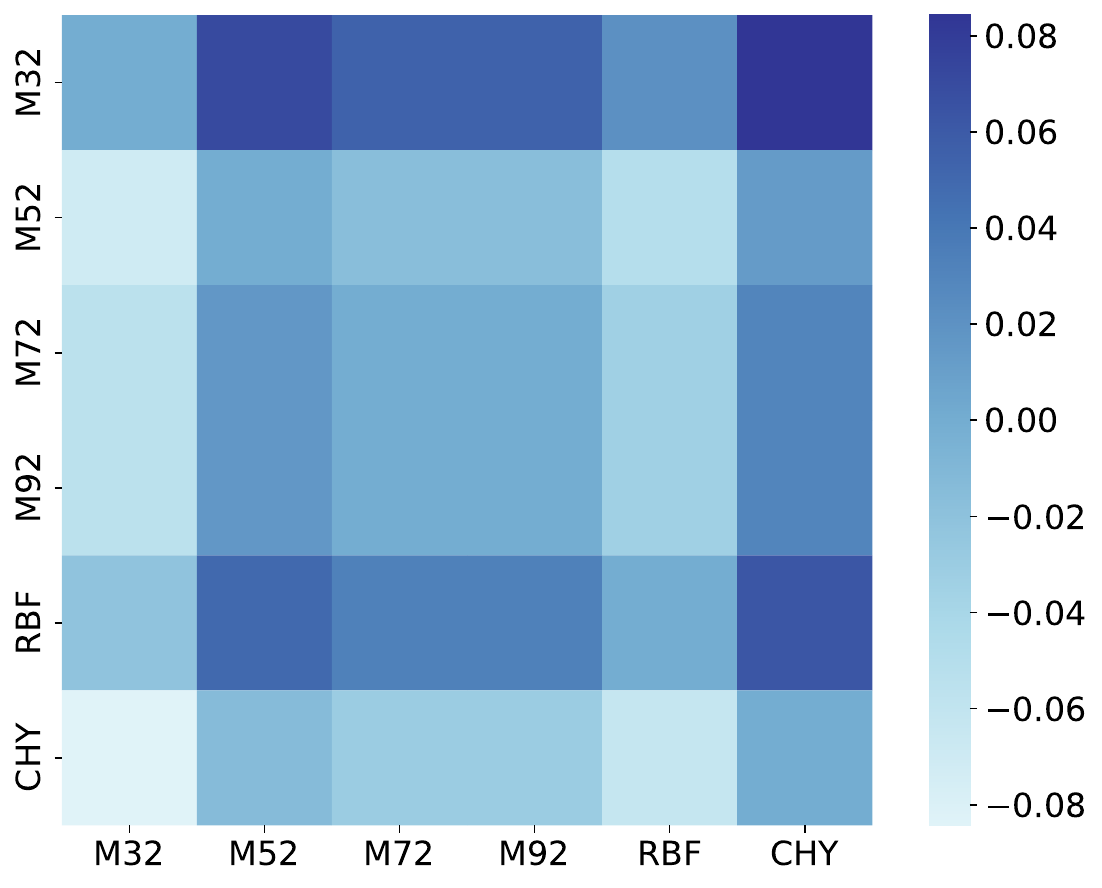}
        \caption{Diag-log}
        \label{fig-ns-sub4}
    \end{subfigure}
    
    \caption{\rep{\textbf{In nested sampling: }Bayes factor heatmaps comparing all pairs of kernel functions within each of the four reconstruction models, based on CC data. Each panel corresponds to one model, and the color scale indicates the strength of preference according to the Bayes factor. Heatmap values correspond to evidence ratios (horizontal/vertical). }{Bayes factor heatmaps comparing kernel pairs across the four modeling frameworks in NS.}}
    \label{fig:ns_heatmap}
\end{figure}

\begin{figure}[t]
    \centering
    \noindent % 取消缩进（可选）
    % 第一行
    \begin{subfigure}[b]{0.42\textwidth} % 减小宽度
        \includegraphics[width=\linewidth]{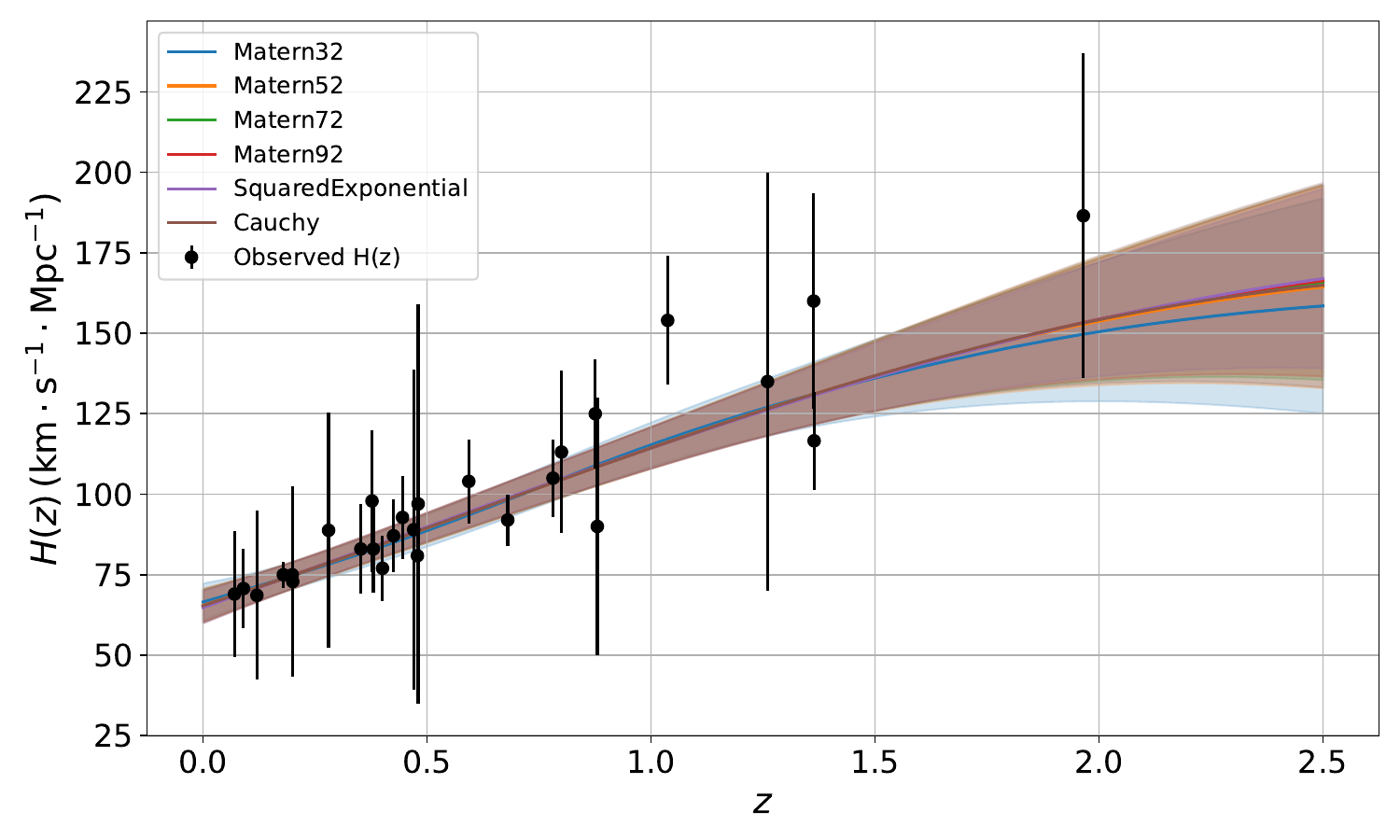}
        \caption{Full-z}
        \label{fig-ns-re1}
    \end{subfigure}
    \hfill
    \begin{subfigure}[b]{0.42\textwidth}
        \includegraphics[width=\linewidth]{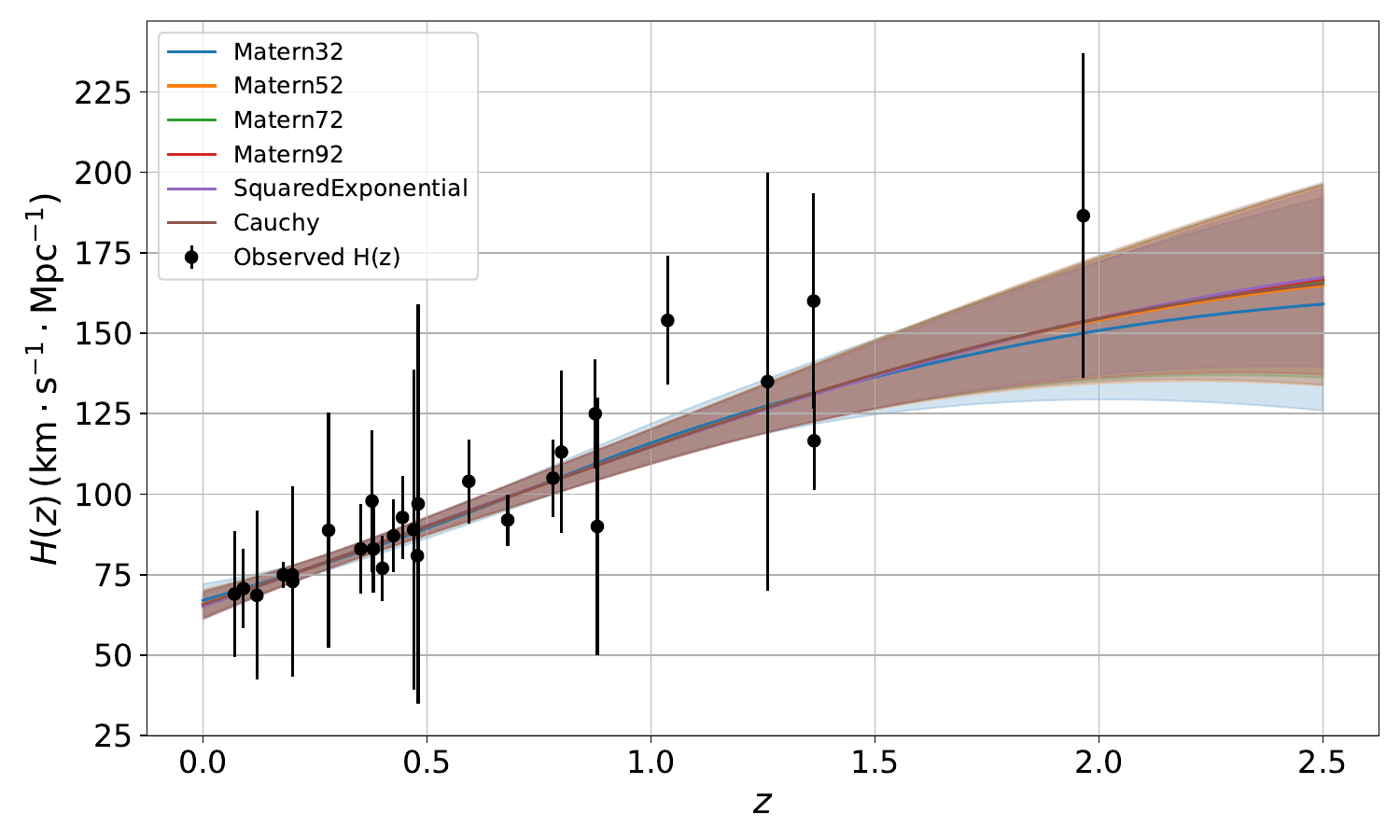}
        \caption{Diag-z}
        \label{fig-ns-re2}
    \end{subfigure}
    
    % 第二行
    \vspace{0.5em}
    \begin{subfigure}[b]{0.42\textwidth}
        \includegraphics[width=\linewidth]{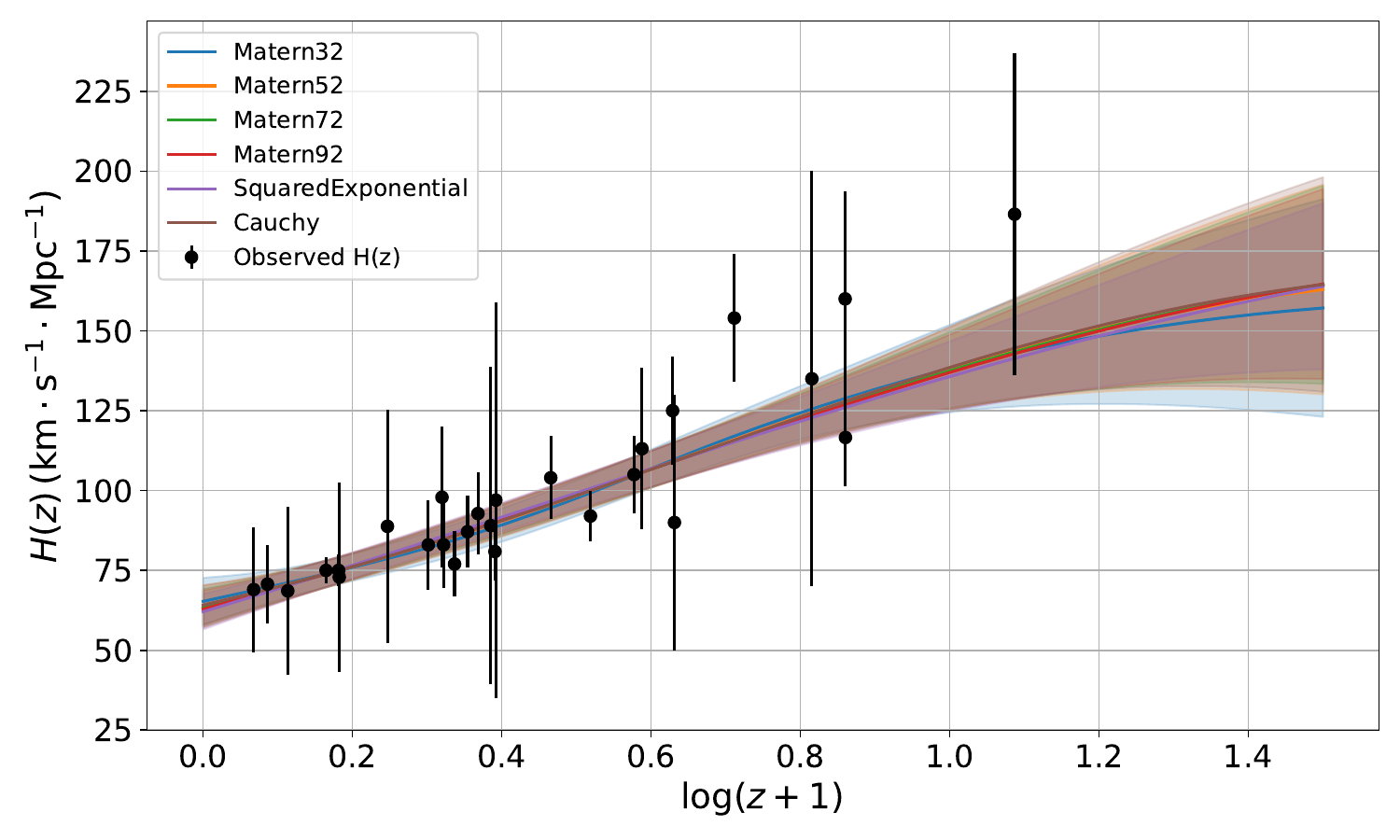}
        \caption{Full-log}
        \label{fig-ns-re3}
    \end{subfigure}
    \hfill
    \begin{subfigure}[b]{0.42\textwidth}
        \includegraphics[width=\linewidth]{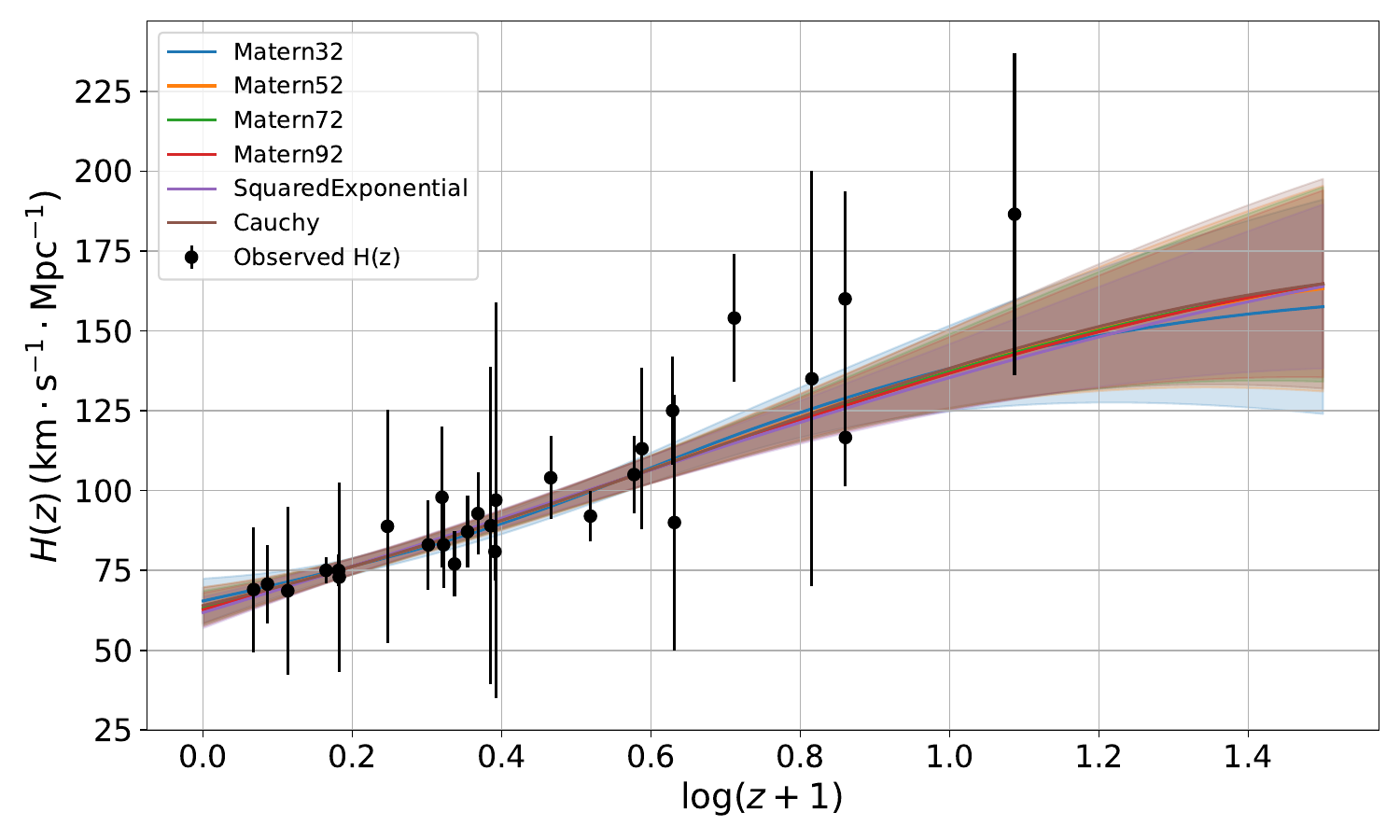}
        \caption{Diag-log}
        \label{fig-ns-re4}
    \end{subfigure}
    
    \caption{\rep{Reconstructed Hubble parameter $H(z)$ from cosmic CC data using nested sampling. Each panel corresponds to one of the four reconstruction models listed in Table~\ref{tab:model}. Within each panel, six kernel functions are shown, reconstructed under their respective optimal hyperparameters in NS. The shaded bands around each reconstruction denote the 1$\sigma$ uncertainties of the Gaussian process predictions. The CC measurements with 1$\sigma$ error bars are plotted for reference.}{Reconstructed Hubble parameter $H(z)$ under optimal hyperparameters for different kernels across the four models in Table~\ref{tab:model}.}}
    \label{fig:ns-re}
\end{figure}

\begin{figure}[H]
    \centering
    \includegraphics[width=0.6\linewidth]{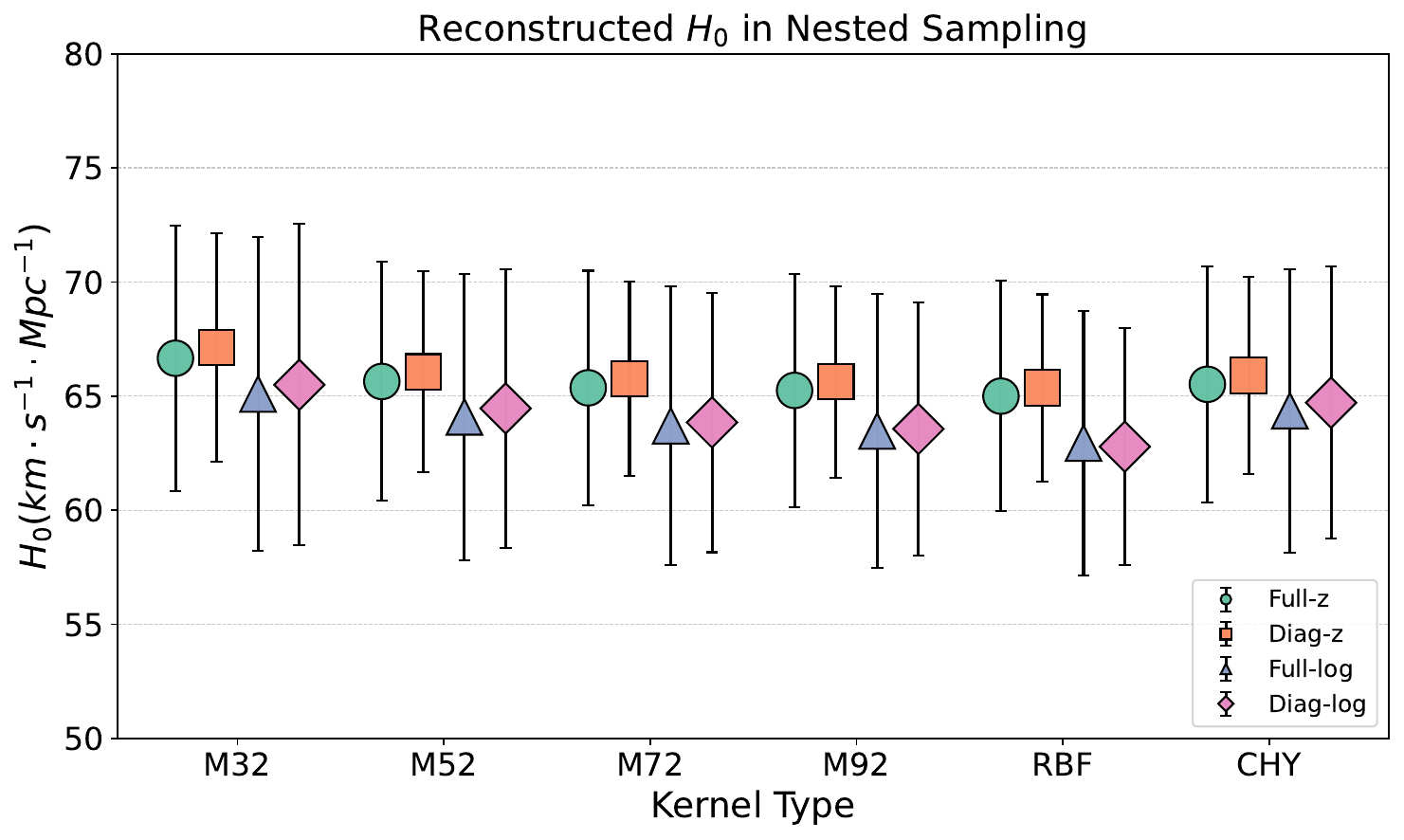}
    \caption{\rep{Comparison of the estimated Hubble constant $H_0$ across different kernel functions and reconstruction models using nested sampling. Different colors indicate the four reconstruction models. The error bars represent the 1$\sigma$ (68\% confidence level) uncertainties from the posterior distributions.}{Comparison of estimated Hubble constant $H_0$ across kernel functions and models. Different colors indicate different models.}}
    \label{fig:nsh0}
\end{figure}

To more clearly demonstrate the differences between kernel functions, Figure~\ref{fig:ns_heatmap} presents Bayes factor heatmaps comparing kernel pairs in four models. The Bayes factor heatmaps in Figure \ref{fig:ns_heatmap} reveal nuanced kernel preference patterns. A comparison of the heat maps between the $z$ space and $log(z+1)$ space reveals that M32 consistently underperforms in both spaces. Interestingly, while the RBF kernel exhibits strong performance in $z$ space, its effectiveness \rep{decreases noticeably}{significantly deteriorates} in $\log(z+1)$ space. Although Table~\ref{tab:evidence_strength} categorizes all kernels as 'Inconclusive' based on traditional evidence thresholds, these heatmaps uncover subtle but \rep{discernible}{significant} preferences. Despite similar \rep{evidence}{LML} values, further analysis in Figure~\ref{fig:nsh0} reveals that seemingly negligible differences in evidence can lead to pronounced divergences in the reconstruction of Hubble constant $H_0$.

In NS analysis, we extract the optimal hyperparameters for each kernel across all models and apply them to the reconstruction process. The results, presented in Figure~\ref{fig:ns-re}, reveal that the M32 kernel — despite its comparatively low evidence values in Figure~\ref{fig:ns} — produces markedly different reconstruction outcomes compared to other kernels in all four models. This finding supports our claim that even marginal differences in evidence can translate into substantial variations in reconstruction quality, underscoring the importance of deliberate kernel selection.

To quantitatively evaluate the reconstruction results, we compute the Hubble constant $H_0$ using posterior samples shown in Figure~\ref{fig:nsh0}. Reconstructions in $z$ space consistently yield higher $H_0$ values than those in $\log(z+1)$ space, with the magnitude of difference being approximately $1.6 \,\text{km}\cdot \text{s}^{-1} \cdot \text{Mpc}^{-1}$, in agreement with \rep{evidence}{LML} trends in Figure~\ref{fig:ns}. Despite minimal differences in \rep{evidence}{LML}, $H_0$ estimates vary \rep{noticeably}{significantly} across kernels, underscoring the magnification effect in parameter inference. The amplification effect highlights critical implications for $H_0$ tension studies, as minor methodological choices can introduce \rep{appreciable}{significant} biases in cosmological conclusions. \rep{This amplification effect has critical implications for $H_0$ tension studies. Across the four reconstruction models for the CC data, the reconstructed Hubble constant $H_0$ shows a kernel-induced spread of $\Delta H_0 \simeq 1.7$–2.7 km·s$^{-1}$·Mpc$^{-1}$, while the typical statistical uncertainty from the posterior samples is $\langle \sigma_{H0} \rangle \simeq 4$–6 km·s$^{-1}$·Mpc$^{-1}$. This corresponds to a relative deviation of roughly 0.3–0.4$\sigma$ across all reconstructions. Although subdominant compared to the statistical errors, the absolute magnitude of the kernel-induced variation is non-negligible, reaching up to $\sim 2.7$ km·s$^{-1}$·Mpc$^{-1}$, representing nearly half of the current SH0ES-Planck tension ($5.6 ~\text{km}\cdot \text{s}^{-1} \cdot \text{Mpc}^{-1}$)} {This amplification effect has critical implications for $H_0$ tension studies: the reconstructed $H_0$ values vary by up to $\Delta H_0 \approx 2.7~\text{km}\cdot \text{s}^{-1} \cdot \text{Mpc}^{-1}$ across kernels, representing nearly half of the current SH0ES vs. Planck tension ($5.6 ~\text{km}\cdot \text{s}^{-1} \cdot \text{Mpc}^{-1}$)} \citep{riess2022comprehensive,2020A&A...641A...6P}.

The reconstruction results of the kernels M32 and RBF have behaviors different from those of the other kernels. The M32 kernel, in particular, not only exhibits the lowest evidence but also yields the largest $H_0$ uncertainties. This aligns with its anomalous behavior in Figure~\ref{fig:ns-re}, supporting our conclusion about the critical importance of kernel selection. Furthermore, full covariance models produce broader uncertainty intervals, suggesting that increased error propagation in the covariance matrix may compromise the overall reconstruction accuracy. The RBF kernel also exhibits interesting behavior as in Figure~\ref{fig:ns_heatmap}. While performing best in the $z$ space, its $\log(z+1)$ transformation results show deviant $H_0$ estimates (Figure \ref{fig:nsh0}). This highlights the nontrivial interaction between kernel function choice and input space transformation.

In summary, our NS analysis reveals that the choice of kernel function plays a pivotal role in GP-based cosmological inference. Although $\ln\mathcal{Z}$ comparisons may suggest only minor differences, these are \rep{numerically}{significantly} amplified in the resulting reconstructions and parameter estimates. Consequently, rigorous kernel selection is essential, even when Bayesian factor alone appears 'Inconclusive'.

\subsection{Results from ABC rejection \add{with CC data}}
\begin{figure}[H]
    \centering
    \includegraphics[width=0.6\linewidth]{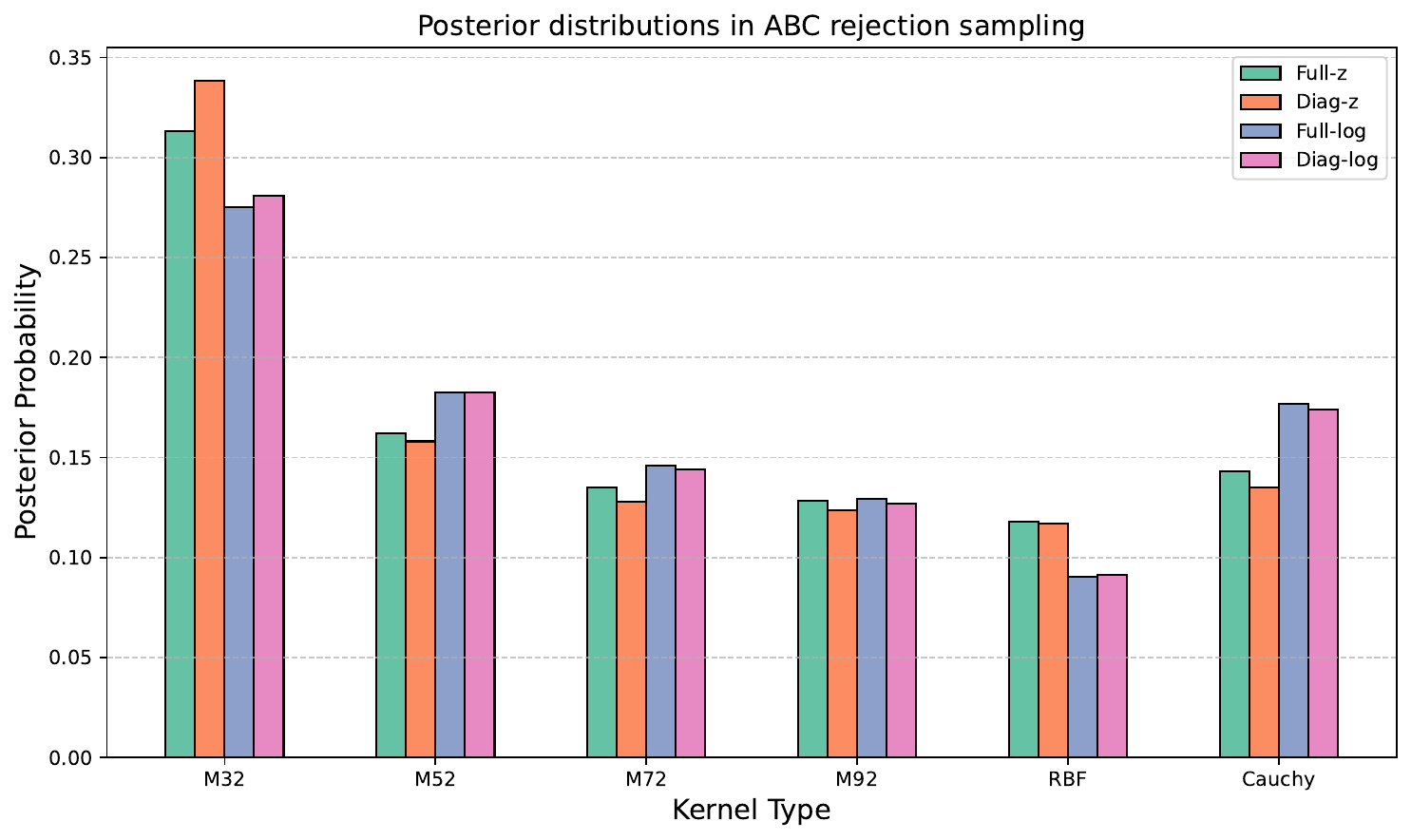}
    \caption{\rep{Normalized posterior probabilities of different kernel functions obtained from the ABC rejection analysis of the CC data, shown across the four reconstruction models in Table~\ref{tab:model}. Colors indicate different models; the vertical axis shows their corresponding normalized probabilities. These values reflect the relative support for each kernel within a given model after normalization.}{Final normalized probabilities of different kernel functions across the four models. Colors represent different models; the horizontal axis denotes kernel types; and the vertical axis shows the normalized probabilities in each model.}}
    \label{fig:abc}
\end{figure}

Our ABC rejection analysis reveals markedly different results compared to NS method. It can be seen from Figure \ref{fig:abc} that the M32 kernel, which performs the worst in NS, emerges as the top performer in ABC rejection. This striking reversal underscores fundamental methodological differences between the two approaches. The NS method, being likelihood-based, maximizes \rep{evidence}{LML}, which inherently penalizes model complexity through Occam's razor, thereby favoring smoother reconstructions. In contrast, the ABC rejection method, which is $\chi^2$-based in our work, minimizes the distance between simulated and observed data, thus prioritizing precise data matching. From a cosmological perspective, the choice between these methods depends on the specific scientific goal. For instance, when the objective is model comparison, such as testing different dark energy models, the NS method with \rep{evidence}{LML} is theoretically more rigorous because it integrates over the entire parameter space. Conversely, for tasks focused on data reconstruction, such as deriving trends in $H(z)$, the ABC method with $\chi^2$ might be more suitable since it directly optimizes the ABC rejection distance for empirical agreement. The superior performance of the M32 kernel in the ABC framework suggests that its parametric form aligns well with the error structures inherent in CC data. However, its poor performance under the NS method warns against overinterpreting these results without considering the Bayesian evidence that supports them.

Our ABC rejection analysis yields strikingly divergent results compared to the NS approach. As evident in Figure~\ref{fig:abc}, the M32 kernel - which demonstrates the poorest performance in NS - emerges as the best performing kernel in the ABC rejection framework.

\begin{figure}[t]
    \centering
    \noindent % 取消缩进（可选）
    % 第一行
    \begin{subfigure}[b]{0.42\textwidth} % 减小宽度
        \includegraphics[width=\linewidth]{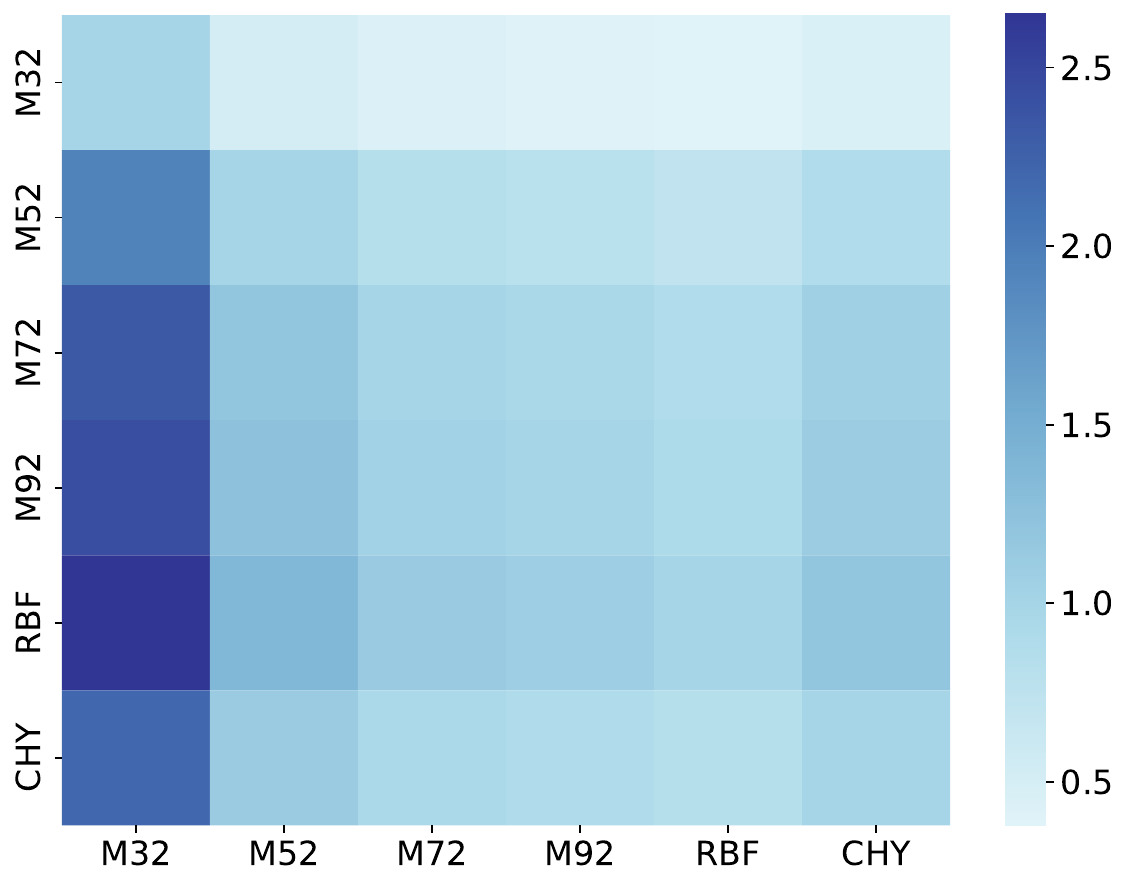}
        \caption{Full-z}
        \label{fig:sub1}
    \end{subfigure}
    \hfill
    \begin{subfigure}[b]{0.42\textwidth}
        \includegraphics[width=\linewidth]{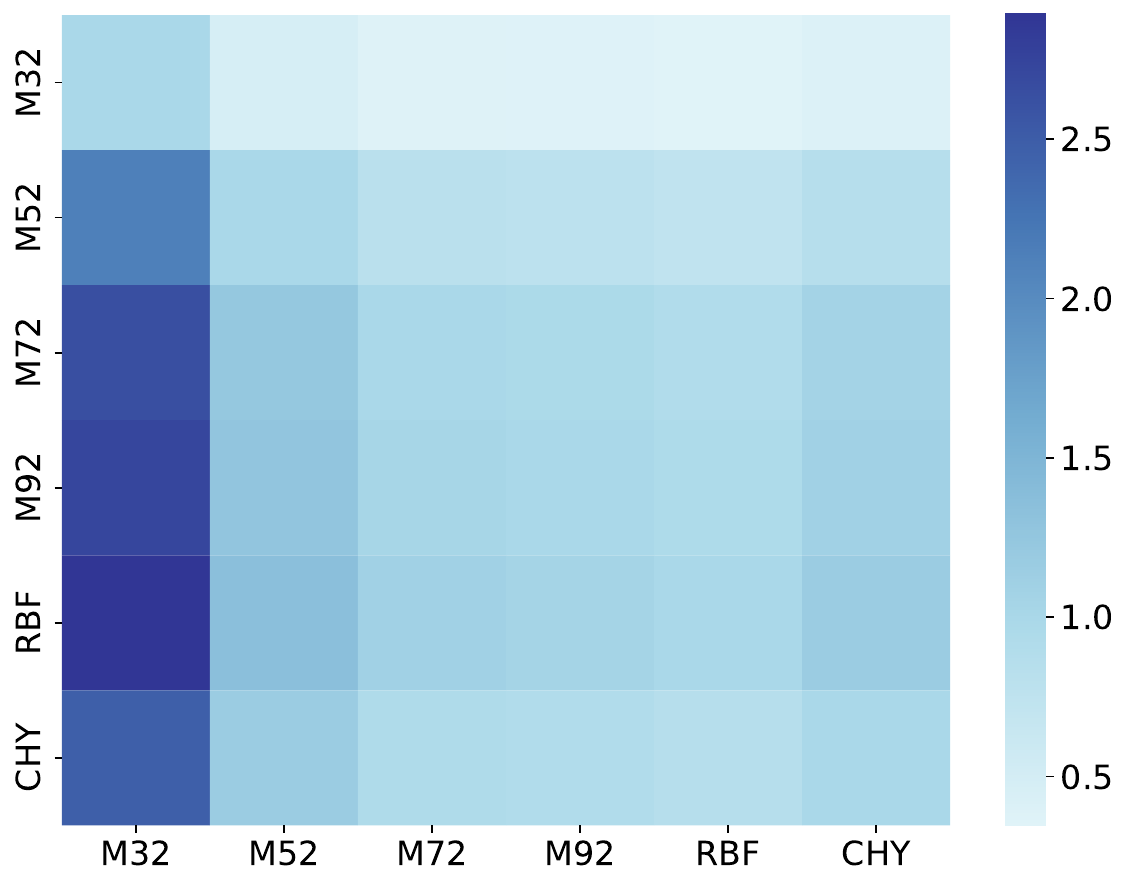}
        \caption{Diag-z}
        \label{fig:sub2}
    \end{subfigure}
    
    % 第二行
    \vspace{0.5em}
    \begin{subfigure}[b]{0.42\textwidth}
        \includegraphics[width=\linewidth]{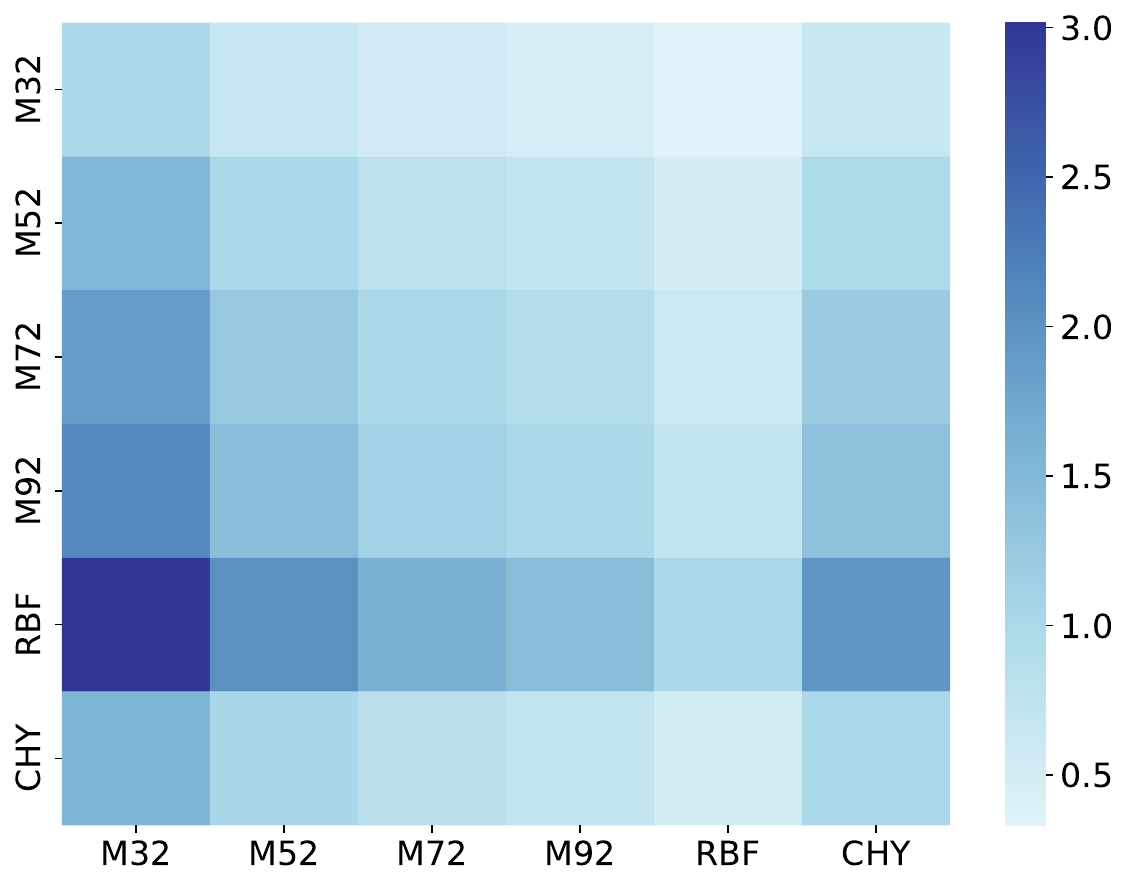}
        \caption{Full-log}
        \label{fig:sub3}
    \end{subfigure}
    \hfill
    \begin{subfigure}[b]{0.42\textwidth}
        \includegraphics[width=\linewidth]{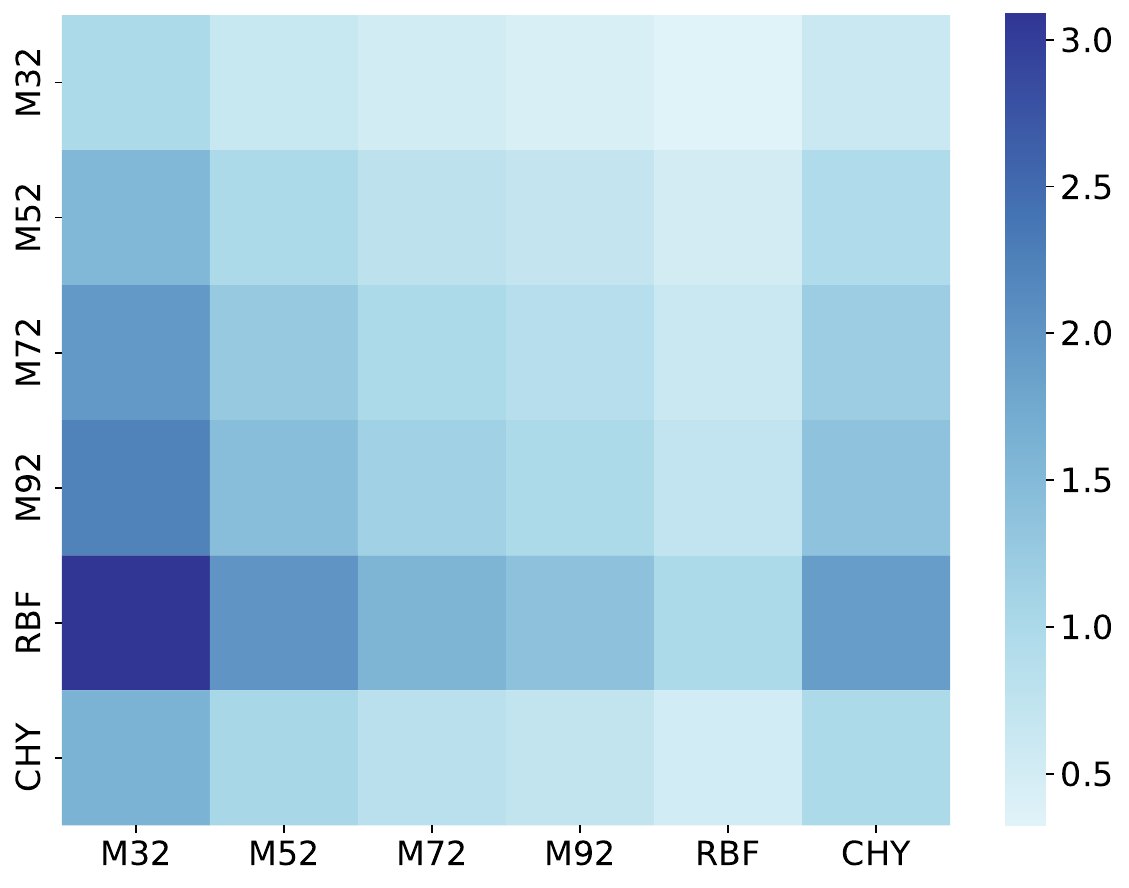}
        \caption{Diag-log}
        \label{fig:sub4}
    \end{subfigure}
    
    \caption{\rep{\textbf{In ABC rejection: }Heatmaps of unnormalized probabilities comparing kernel pairs across the four modeling frameworks. The color scale indicates the relative probability values (horizontal/vertical) prior to normalization.}{Heatmaps comparing kernel pairs across the four modeling frameworks in ABC rejection.}}
    \label{fig:abc4}
\end{figure}

The comparison of Bayes factors in Figure~\ref{fig:abc4} reveals that M32 ranks as the top-performing kernel among the four models, while RBF performs the worst. However, we observe that even the best-performing M32 kernel only achieves a 'Barely worth mention' classification in Table~\ref{tab:bayes_grades} when compared to the lowest-ranked RBF kernel. This indicates that the performance difference between them is statistically insignificant, a finding that aligns with the results shown in Section~\ref{sec:nested_results}.

\begin{figure}[t]
    \centering
    \includegraphics[width=0.75\linewidth]{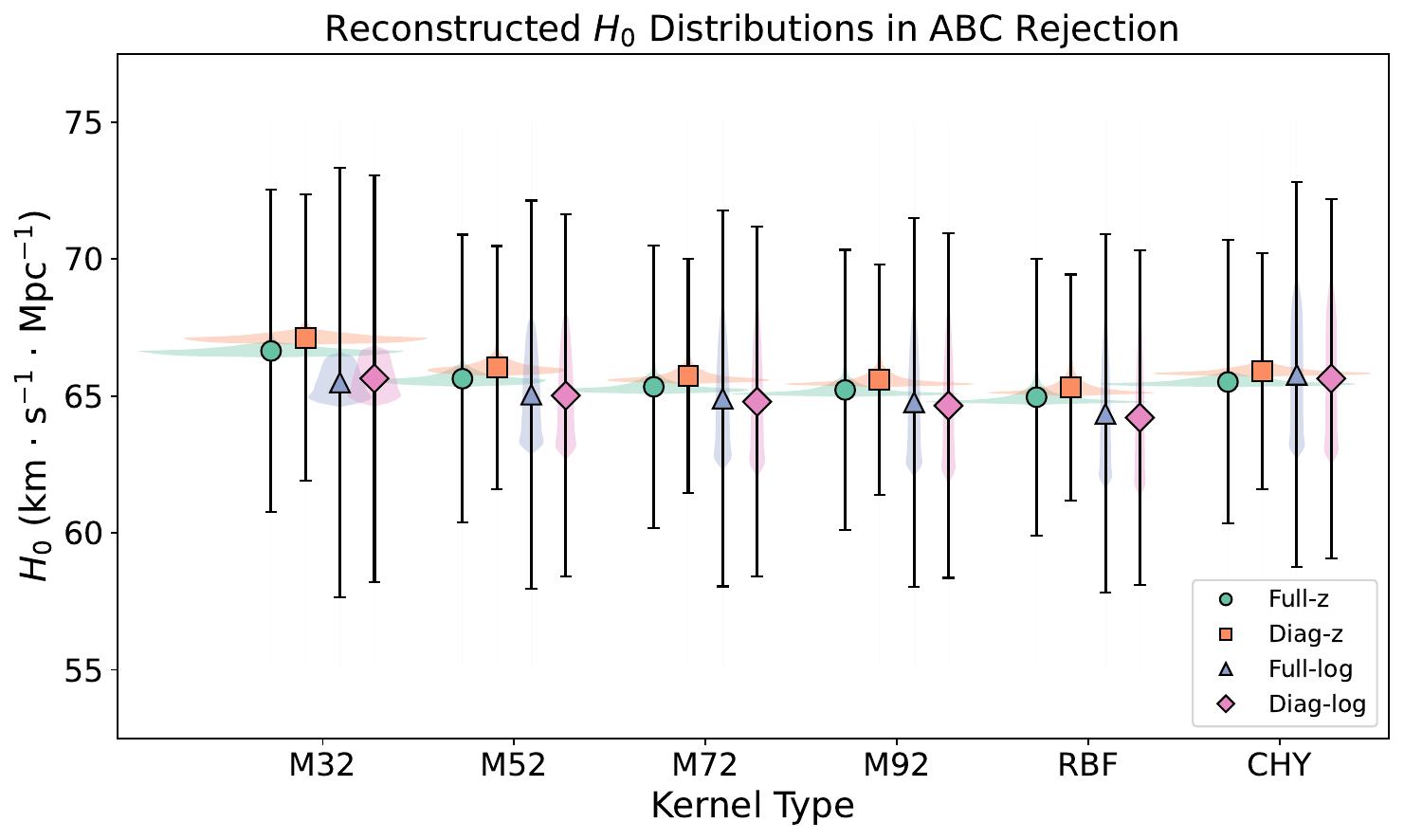}
    \caption{\rep{Hubble constant estimates ($H_0$) for different kernel functions across the four reconstruction models using ABC rejection. Colors represent different models. The error bars denote the 1$\sigma$ uncertainties derived from the ABC posterior samples. In addition, semi-transparent shaded areas represent the kernel density estimates (KDEs) of the $H_0$ posterior distributions, providing a visualization of the full distribution beyond the mean and error bars.}{Hubble constant estimates ($H_0$) for different kernel functions across the four models. Colors represent different models; the horizontal axis indicates the kernel type; and the vertical axis corresponds to the estimated $H_0$ values}}
    \label{fig:abch0}
\end{figure}

To investigate these differences more thoroughly, we examine the Hubble constant reconstruction across all four models, as shown in Figure~\ref{fig:abch0}. The figure also incorporates the sampling point kernel density estimation (KDE) distributions obtained through the Sampling point \citep{chen2017tutorial}. These distributions are probabilistically consistent with the results presented in Figure~\ref{fig:abc}. 
\add{The four reconstruction methods for CC data yield a Hubble constant $H_0$ with a kernel-driven variation of $\Delta H_0 \approx 2.0$–2.9 km·s$^{-1}$·Mpc$^{-1}$. The average statistical uncertainty from posterior samples ranges from $\langle \sigma_{H0} \rangle \approx 4.1$–7.2 km·s$^{-1}$·Mpc$^{-1}$, corresponding to a relative deviation of about 0.3–0.4$\sigma$ across all methods. While the kernel-induced spread is smaller than the statistical uncertainties, its magnitude, up to $\sim 2.9$ km·s$^{-1}$·Mpc$^{-1}$, is substantial, constituting a meaningful portion of the SH0ES–Planck $H_0$ tension, which is consistent with that in nested sampling. This suggests that careful selection of the kernel is important when interpreting $H_0$ reconstructions using Gaussian processes in ABC rejection analysis.} Besides, our analysis reveals that Hubble constant values exhibit tighter clustering in $z$ space compared to the more dispersed distribution observed in $\log(z+1)$ space. However, it should be noted that despite showing the most concentrated distribution, the M32 kernel exhibits the largest estimation errors among all models. \add{Furthermore, in both the nested sampling and ABC rejection analyses, we observe a systematic trend across the Matérn kernels: as the smoothness parameter $\nu$ increases, the associated uncertainty in $H_{0}$ decreases. This behavior is consistent with the findings reported by E. Ó Colgáin et al.~\cite{o2021elucidating}.}

A critical consideration in interpreting these results is the well-documented sensitivity of ABC rejection outcomes to the selection of distance functions \citep{zhangKernelSelectionGaussian2023}. Varying kernel functions can produce substantially divergent results depending on the choice of the distance metric \citep{sisson2018handbook}. 

\subsection{\add{Results with BAO Data}}

\begin{figure}[H]
    \centering
    % 第一行（两张图）
    \begin{subfigure}[b]{0.45\textwidth}
        \includegraphics[width=\linewidth]{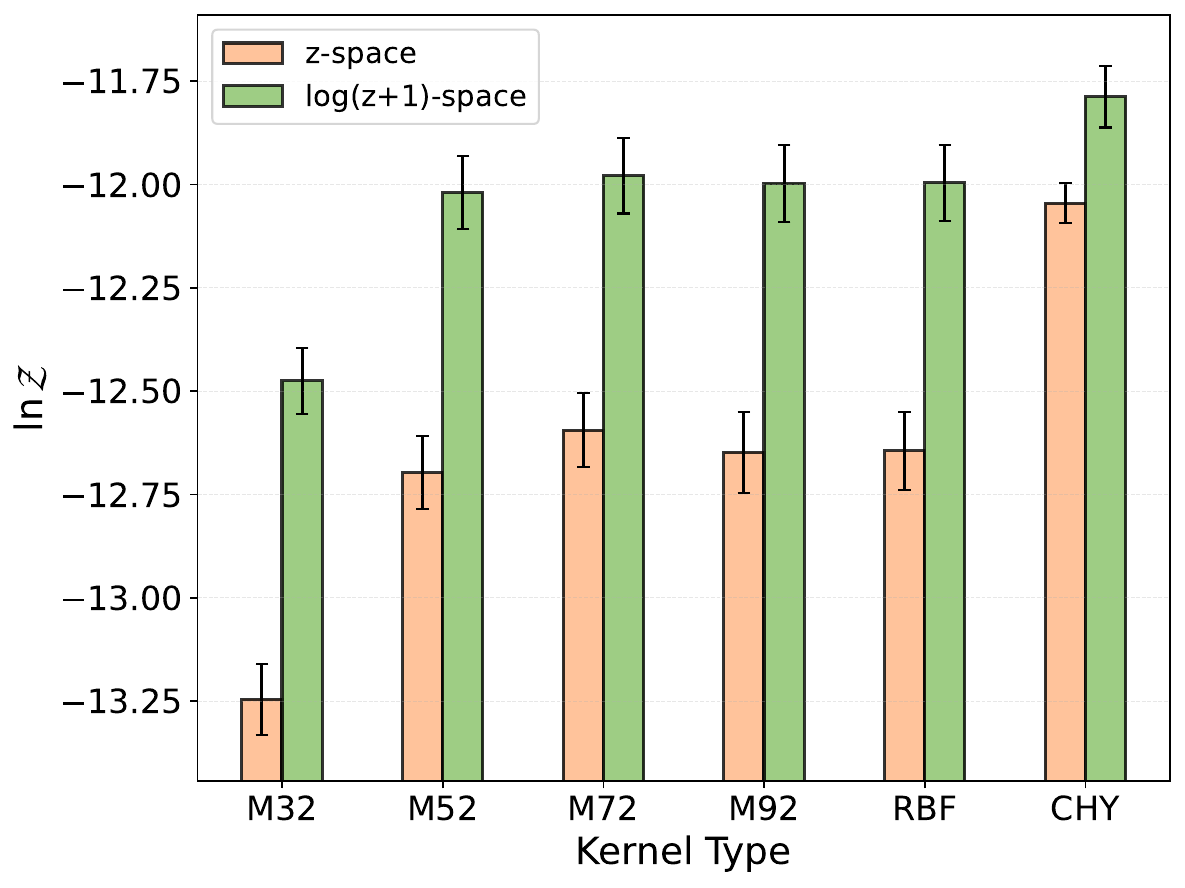}
        \caption{\add{Evidence with uncertainties across kernel functions in nested sampling.}}
        \label{fig:sub1bao}
    \end{subfigure}
    \hfill
    \begin{subfigure}[b]{0.45\textwidth}
        \includegraphics[width=\linewidth]{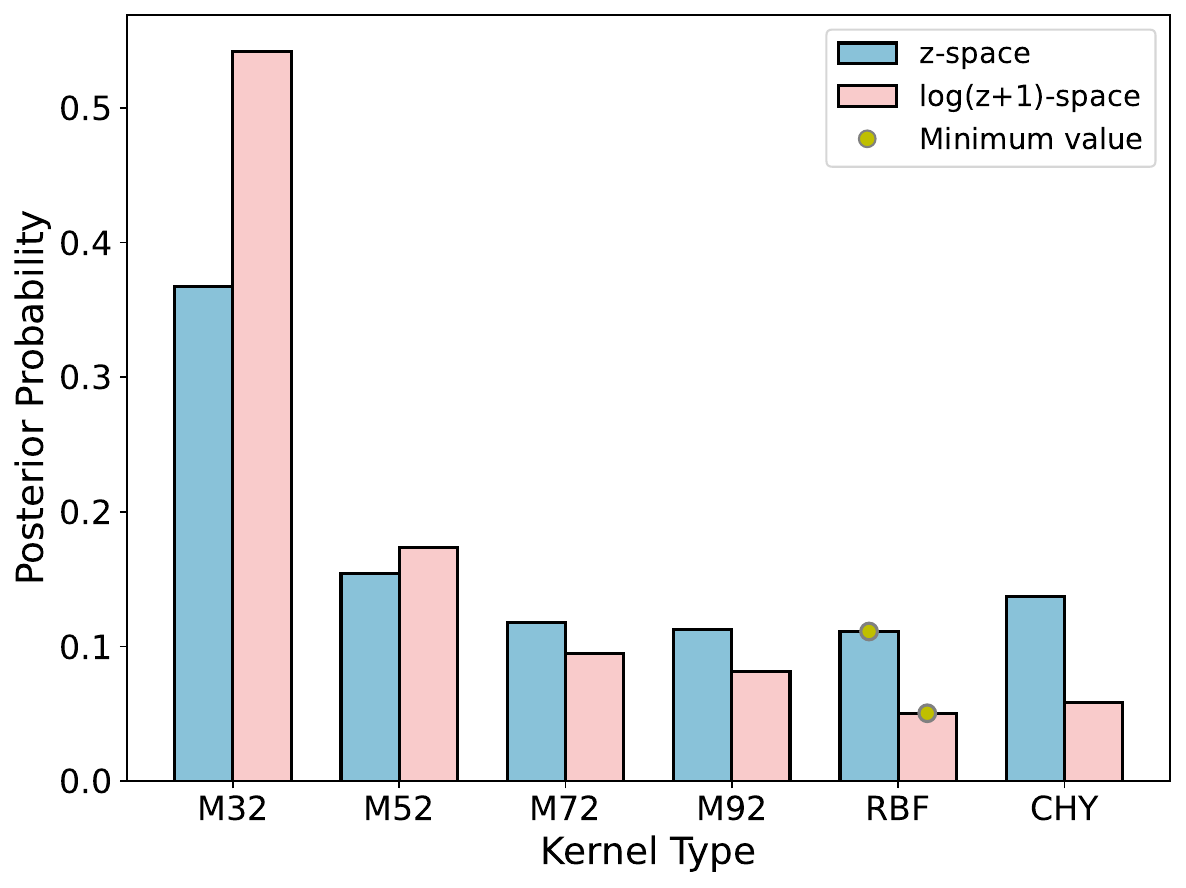}
        \caption{\add{Normalized posterior probabilities of different kernel functions in ABC rejection.}}
        \label{fig:sub2bao}
    \end{subfigure}
    
    \caption{\add{\textbf{BAO Data:} Comparison of kernel function performance under diagonal covariance reconstruction in $z$ and $\log(z+1)$ spaces. Panel (a) shows evidence and associated uncertainties from nested sampling, while panel (b) presents normalized probabilities obtained via ABC rejection.}}
    \label{fig:bao}
\end{figure}

\add{We apply the BAO data to the GP for the same type of analysis, with the results presented in Figure \ref{fig:bao}. Figure \ref{fig:sub1bao} shows the nested sampling outcomes. We find that in both parameter spaces, the evidence differences between kernels remain small ($\Delta \ln\mathcal{Z} < 1.25$), consistent with the results from the CC data. This indicates that the kernels are largely comparable, with the four central kernel functions exhibiting substantial overlap within $1\sigma$. In contrast, the Matérn32 and Cauchy kernels deviate more noticeably from the central group. Unlike the CC case, however, the BAO data favors the $\log(z+1)$ space: within each kernel, the evidence in $\log(z+1)$ is consistently higher than in $z$ space. Figure \ref{fig:sub2bao} presents the ABC rejection results. Here we again find that the BAO behavior broadly agrees with the CC data in Figure \ref{fig:abc}. Notably, the Matérn32 kernel, which performed the worst under nested sampling, emerges as the best-performing model in both spaces, whereas the RBF kernel yields the poorest performance.}

\subsection{\add{Results with SN Ia Data}}

\begin{figure}[H]
    \centering
    % 第一行（两张图）
    \begin{subfigure}[b]{0.45\textwidth}
        \includegraphics[width=\linewidth]{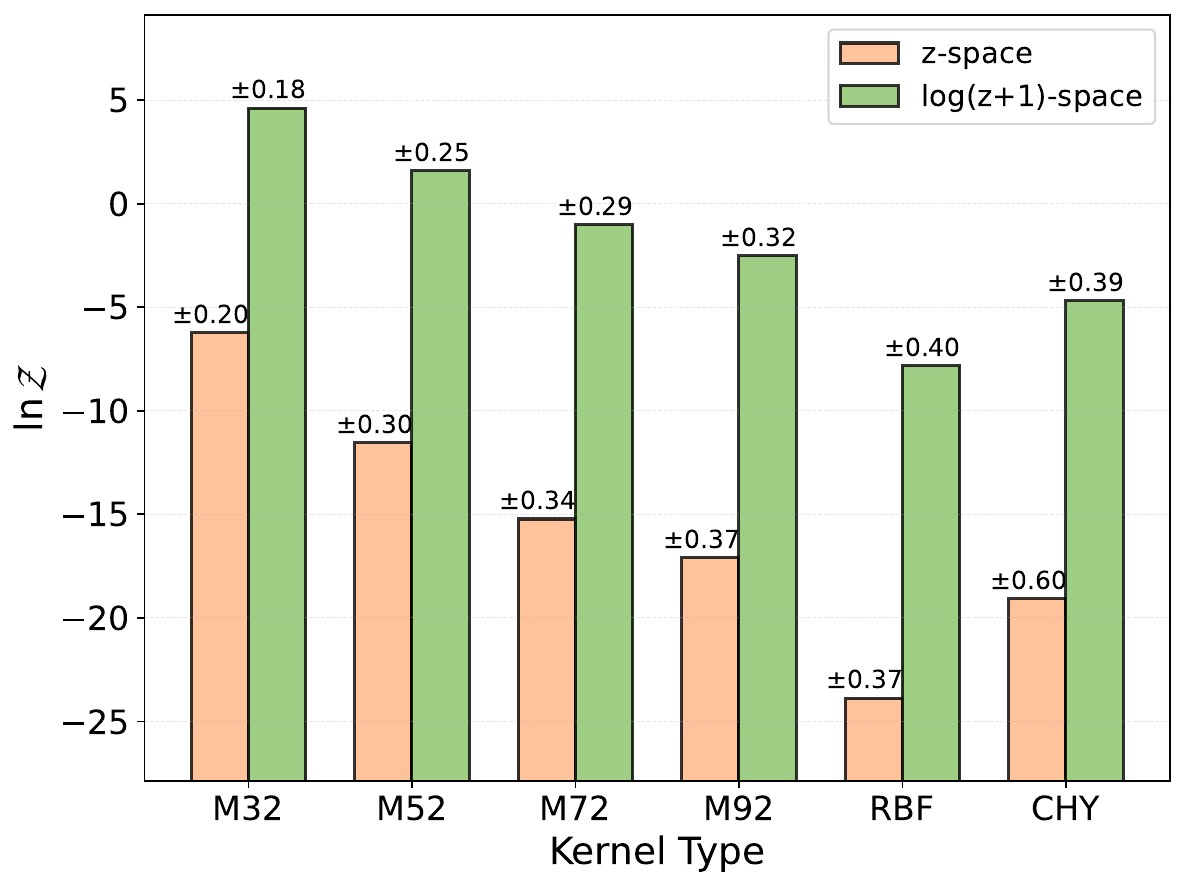}
        \caption{\add{Evidence with uncertainties across kernel functions in nested sampling.}}
        \label{fig:sub1sn}
    \end{subfigure}
    \hfill
    \begin{subfigure}[b]{0.45\textwidth}
        \includegraphics[width=\linewidth]{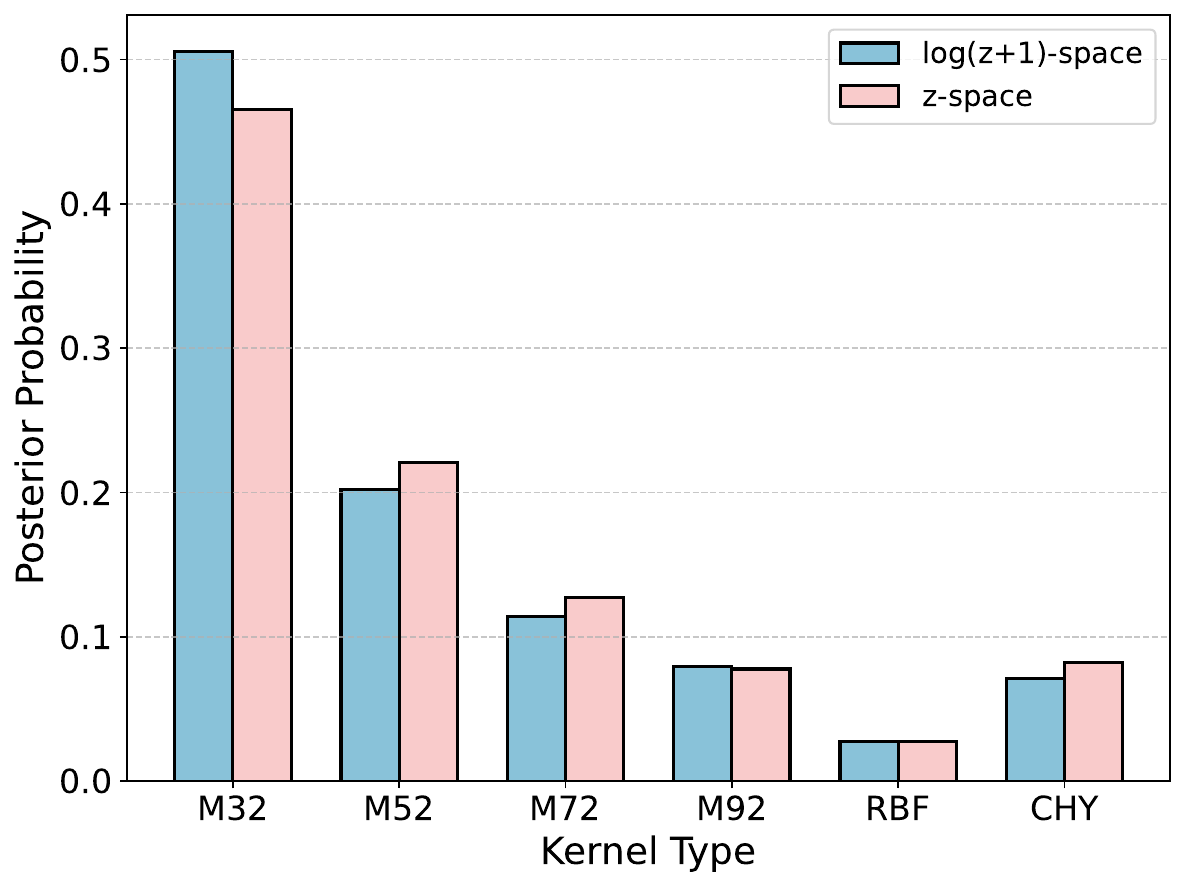}
        \caption{\add{Normalized posterior probabilities of different kernel functions in ABC rejection.}}
        \label{fig:sub2sn}
    \end{subfigure}
    
    \caption{\add{\textbf{SN Ia Data:} Comparison of kernel function performance  under full covariance reconstruction in $z$ and $\log(z+1)$ spaces. Panel (a) shows evidence and associated uncertainties from nested sampling, while panel (b) presents normalized probabilities obtained via ABC rejection.}}
    \label{fig:sn}
\end{figure}

\add{After reconstructing the SN Ia data with GP, we obtain results that differ markedly from those of the CC and BAO datasets, as shown in Figure~\ref{fig:sn}. In Figure~\ref{fig:sub1sn}, which displays the nested sampling outcomes, the differences among kernel functions become pronounced. Interestingly, the Matérn32 kernel, previously the poorest performer with CC and BAO, now provides the best fit in both parameter spaces. Furthermore, the SN Ia results show higher evidence in the $\log(z+1)$ space than in the $z$ space, consistent with the BAO findings. This indicates that, in the case of the BAO and SN Ia datasets, the $\log(z+1)$ space provides a more suitable framework for reconstruction than the $z$ space.}

\add{Figure~\ref{fig:sub2sn} shows the ABC rejection results, which are broadly similar to those from the CC and BAO datasets: Matérn32 continues to yield the best performance, while the RBF kernel remains the weakest. Taken together, the analyses across the three datasets reveal that Matérn32 can sometimes be the strongest and sometimes the weakest performer, which underscores the importance of kernel selection in GP reconstruction.}

\subsection{\add{Joint Analysis of BAO and SN Ia Data with CMB Prior}}
\begin{figure}[H]
    \centering
    % 第一行（两张图）
    \begin{subfigure}[b]{0.45\textwidth}
        \includegraphics[width=\linewidth]{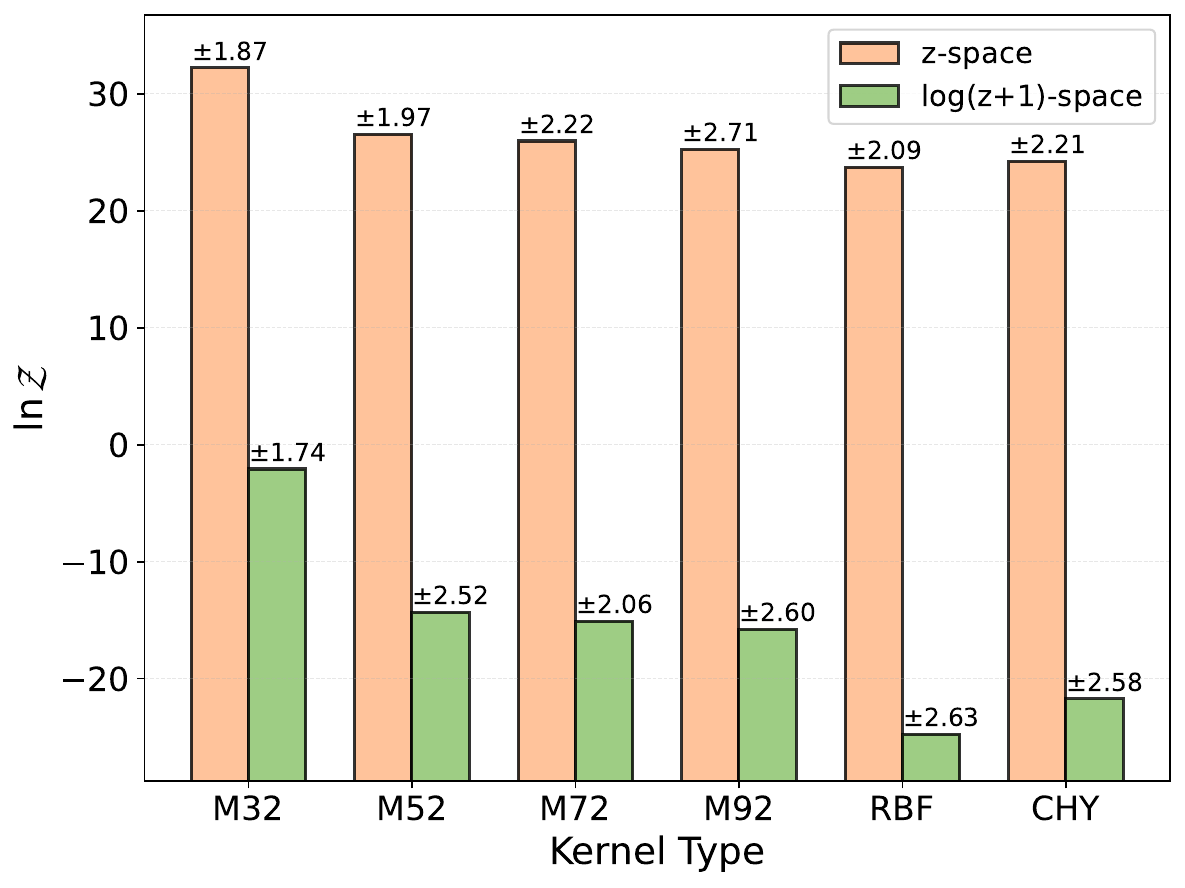}
        \caption{\add{Evidence with uncertainties across kernel functions in nested sampling.}}
        \label{fig:sub1joint}
    \end{subfigure}
    \hfill
    \begin{subfigure}[b]{0.45\textwidth}
        \includegraphics[width=\linewidth]{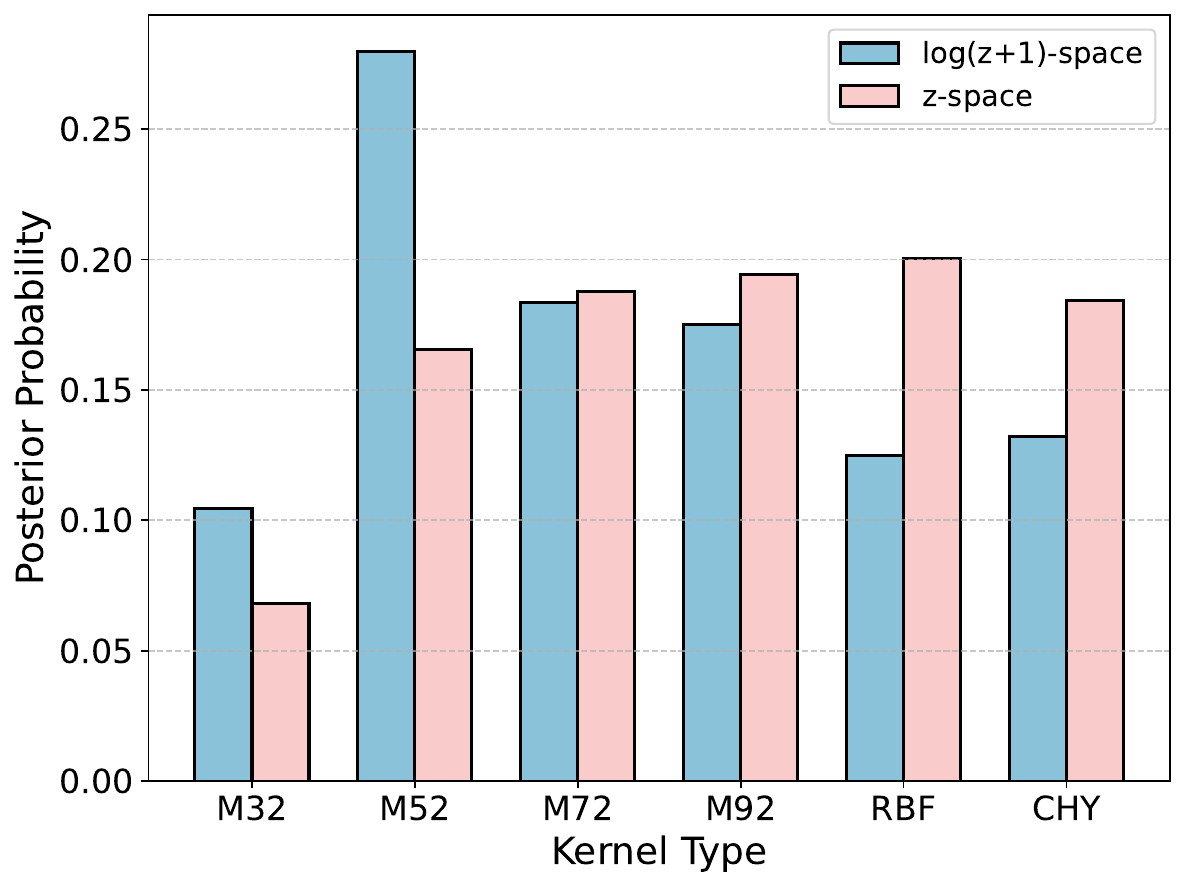}
        \caption{\add{Normalized posterior probabilities of different kernel functions in ABC rejection.}}
        \label{fig:sub2joint}
    \end{subfigure}
    
    \caption{\add{\textbf{Joint BAO + SN Ia + CMB Prior:} Comparison of kernel function performance under reconstructions in $z$ and $\log(z+1)$ spaces. Panel (a) shows evidence and associated uncertainties from nested sampling, while panel (b) presents normalized probabilities obtained via ABC rejection.}}
    \label{fig:joint1}
\end{figure}

\add{Combining the CMB prior, we perform a joint analysis of the BAO and SN Ia data. The results from the GP are presented in Figure~\ref{fig:joint1}. Figure~\ref{fig:sub1joint} shows the nested sampling outcomes, which differ markedly from the separate BAO and SN Ia analyses. Here, the reconstruction in the $z$ space is noticeably better than in the $\log(z+1)$ space. After combining the BAO and SN Ia datasets with the CMB prior, the performance in redshift space is reversed relative to the individual cases, suggesting that in the linear $z$ space, the BAO and SN Ia constraints complement each other more effectively. Interestingly, the Matérn32 kernel, previously the weakest performer, now yields the best results. At the same time, the differences among kernels remain consistent with the BAO case, showing pronounced variation. This further highlights the importance of kernel selection in GP. Figure~\ref{fig:sub2joint} presents the ABC rejection results. In contrast to the nested sampling findings, the Matérn32 kernel performs noticeably worse in both spaces, opposite to the trends observed in the separate analyses of BAO and SN Ia data.}

\begin{figure}[H]
    \centering
    % 第一行（两张图）
    \begin{subfigure}[b]{0.49\textwidth}
        \includegraphics[width=\linewidth]{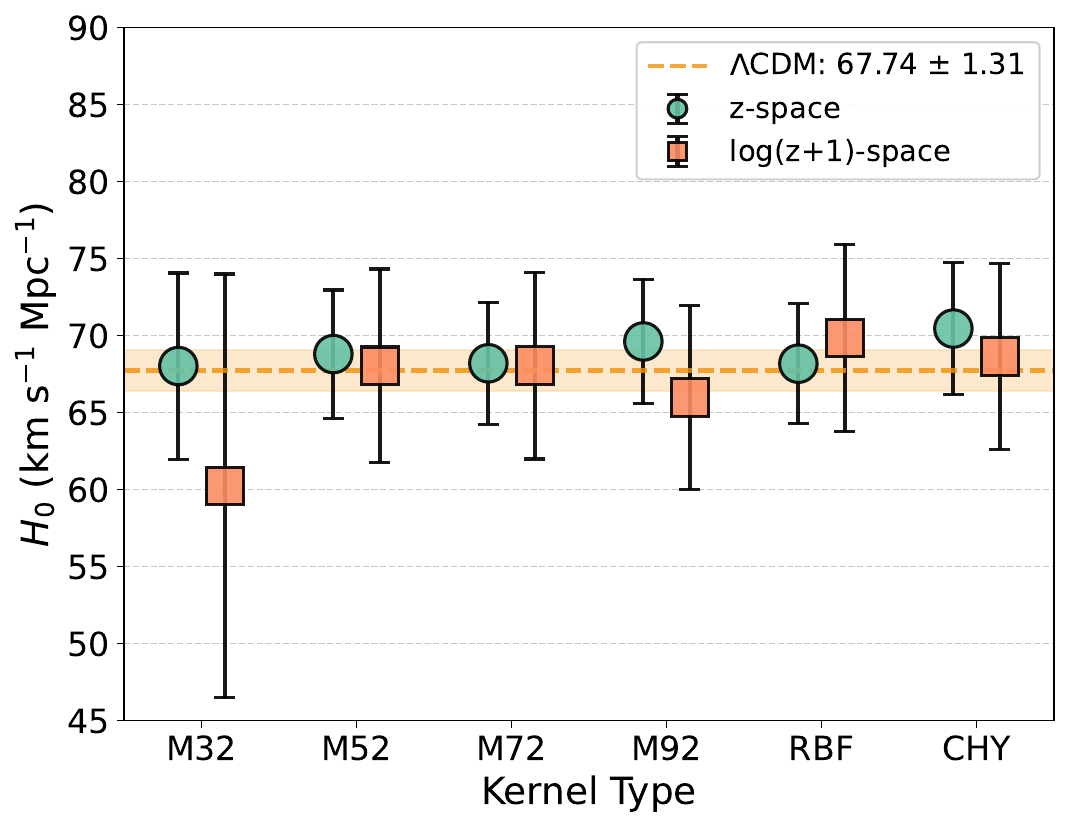}
        \caption{\add{Estimated Hubble constant ($H_0$) with $1\sigma$ uncertainties from nested sampling across different kernel functions. }}
        \label{fig:sub3joint}
    \end{subfigure}
    \hfill
    \begin{subfigure}[b]{0.5\textwidth}
        \includegraphics[width=\linewidth]{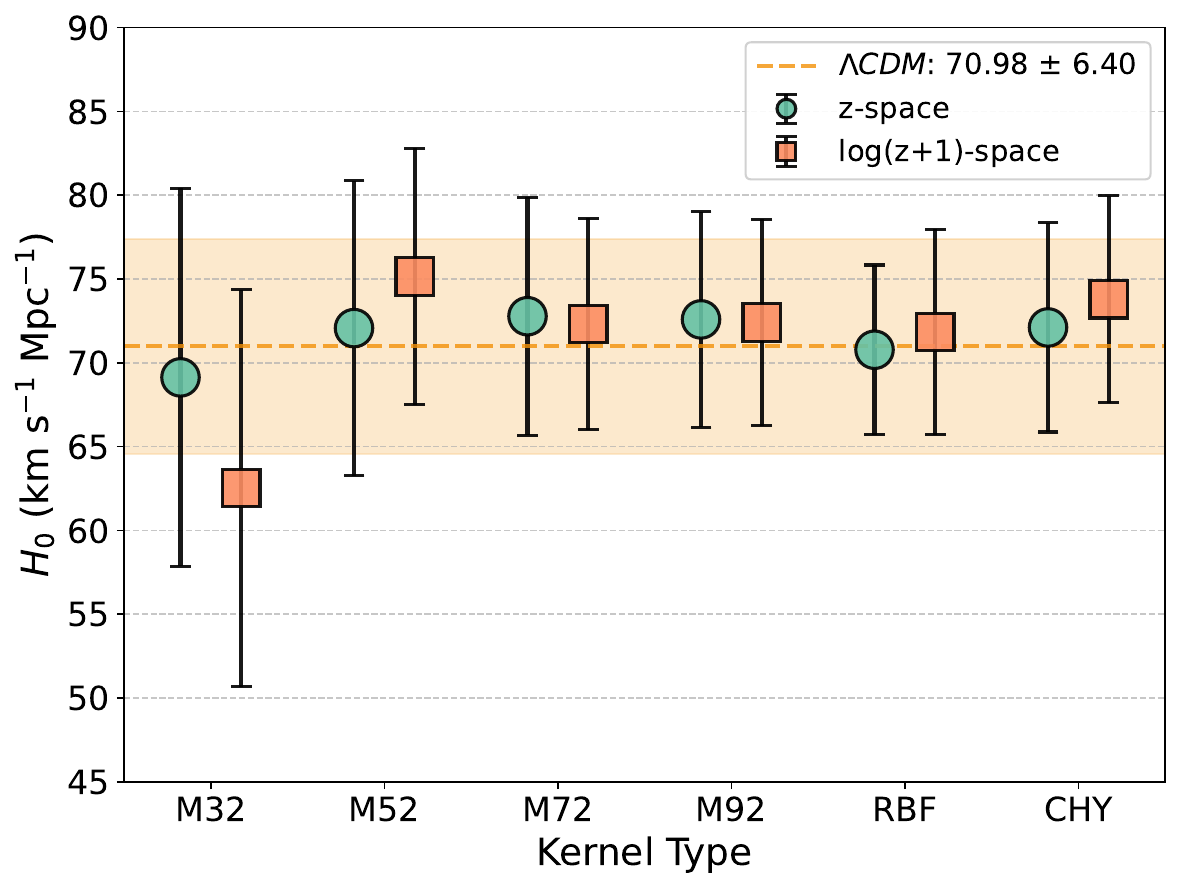}
        \caption{\add{Estimated Hubble constant ($H_0$) with $1\sigma$ uncertainties from ABC rejection across different kernel functions. }}
        \label{fig:sub4joint}
    \end{subfigure}
    
    \caption{\add{Comparison of estimated Hubble constant $H_0$ across kernel functions for the joint BAO + SN Ia data with CMB prior. Results from nested sampling (a) and ABC rejection (b) are shown, with $\Lambda$CDM predictions and uncertainties overlaid for reference.}}
    \label{fig:joint2}
\end{figure}

\add{In the joint analysis of BAO and SN Ia data with the addition of the CMB prior, we find that the two inference approaches lead to different implications for the Hubble tension. For the nested sampling case (Figure~\ref{fig:sub3joint}), the $\Lambda$CDM model yields $H_{0} = 67.74 \pm 1.31$ km s$^{-1}$ Mpc$^{-1}$, and the GP reconstructions cluster around $H_{0} \simeq 68$–70 km s$^{-1}$ Mpc$^{-1}$, closely aligned with the Planck CMB determination and therefore not alleviating the Hubble tension. In this case, the largest deviation arises for the Cauchy kernel in $z$ space ($\Delta H_{0}/\sigma_{\Lambda{\rm CDM}} \simeq 2.1\sigma$), while in $\log(z+1)$ space the Matérn32 kernel shows a particularly large shift ($\Delta H_{0} \simeq 7.5$ km s$^{-1}$ Mpc$^{-1}$, corresponding to $\sim 5.7\sigma$). By contrast, for the ABC rejection case (Figure~\ref{fig:sub4joint}), the $\Lambda$CDM model yields $H_{0} = 70.98 \pm 6.40$ km s$^{-1}$ Mpc$^{-1}$, with GP reconstructions remaining consistent across most kernels (differences below $0.3\sigma$). The only exception is again the Matérn32 kernel in the $\log(z+1)$ space, which deviates by $\sim 8.5$ km s$^{-1}$ Mpc$^{-1}$, but still corresponds to only $\sim 1.3\sigma$ due to the larger $\Lambda$CDM uncertainty in this test case. Importantly, the value obtained with ABC rejection lies closer to the SH0ES distance-ladder determination than to the CMB-inferred value. This contrast highlights how different statistical inference frameworks, as well as kernel and redshift parameterization choices, can shift the interpretation of BAO+SN+CMB prior data with respect to the ongoing Hubble tension.}

\add{Taken together, these results indicate that while the statistical consistency with $\Lambda$CDM is maintained, there are interesting hints of systematic modeling-related deviations, particularly in the $\log(z+1)$ reconstructions with Matérn32. This suggests that kernel selection and redshift parameterization contribute at the level of $\sim 1$–$6\sigma$ relative to the $\Lambda$CDM uncertainty, thereby representing an additional source of systematic uncertainty that should be considered alongside the statistical errors.}

\section{Conclusion} \label{sec:discussion}
In this study, we have systematically explored GP regression for the analysis of CC data, \add{BAO data and SN Ia data}, incorporating the latest methodological advances. \add{We focus on analyzing the situation of CC data under different spatial conditions and different covariance matrices, as well as the situation when BAO and SN Ia are jointly analyzed with CMB prior.} 

\add{In the analysis of CC data,} our framework combines redshift $z$ and $\log(z+1)$ representations, while implementing current best practices in covariance matrix treatment, including recent progress in addressing systematic effects in $D4000$ measurements. This comprehensive approach enables us to fully exploit the information content of increasingly precise low-redshift CC data while properly characterizing uncertainties across the entire redshift range. Our NS analysis demonstrates that the reconstruction in $\log(z+1)$ space is reasonable and the diagonal covariance models consistently achieve superior performance compared to full-covariance implementations. Specifically, the Cauchy kernel yields optimal results in $z$ space, while the RBF kernel shows best performance in $\log(z+1)$ space. In striking contrast, ABC rejection analysis employing $\chi^2$-based distance metrics produces fundamentally different kernel preferences, highlighting the critical dependence of kernel selection on methodological approach. Most notably, the M32 kernel - which demonstrates the poorest performance in NS - emerges as the top-performing choice in the ABC rejection framework. This methodological divergence has profound implications for cosmological parameter inference, particularly for the measurement of $H_0$, where reliance on a single method may introduce systematic biases.  To address these challenges, future studies should explore consensus approaches that integrate insights from multiple methodologies or employ meta-analysis frameworks to synthesize results from diverse analytical techniques.

In both methods, Bayes factor comparisons reveal that the performance advantage of the optimal kernel function over the poorest-performing one is not statistically significant in either model. Nevertheless, our analysis of the final Hubble constant distributions demonstrates that even these subtle differences between kernel functions can lead to substantially divergent sampling outcomes. 
Specifically, the reconstructed $H_0$ values vary by up to $ 2.7~\text{km}\cdot \text{s}^{-1} \cdot \text{Mpc}^{-1}$ across kernels, which is nearly half of the current SH0ES–Planck tension of $5.6~\text{km}\cdot \text{s}^{-1} \cdot \text{Mpc}^{-1}$ \citep{riess2022comprehensive,2020A&A...641A...6P}. This amplification effect highlights that even minor methodological choices in kernel selection can introduce \rep{non-negligible}{significant} biases in cosmological parameter inference. \del{Therefore, kernel choice should not rely solely on marginal performance differences but be guided by a comprehensive analysis aligned with the modeling objectives to ensure robust conclusions.} These findings suggest that kernel selection should not be based solely on marginal differences in performance metrics. Rather, the choice should be guided by comprehensive analysis tailored to the specific objectives of the modeling task, ensuring selection of the most appropriate kernel function for the intended application.

\add{In the joint analysis, our comparison shows that the nested sampling analysis yields $H_{0}$ values consistent with the Planck CMB determination, thereby reinforcing the existing Hubble tension, whereas the ABC rejection analysis produces values closer to the SH0ES distance-ladder result, which reduces the discrepancy relative to SH0ES. In both cases, the Matérn32 kernel in the $\log(z+1)$ space exhibits the largest deviations, underscoring that GP reconstructions are highly sensitive to kernel choice and redshift parameterization. These findings highlight that the interpretation of BAO+SN+CMB prior data with respect to the Hubble tension is not only data-driven but also strongly shaped by the chosen statistical framework.}

Although our initial objective sought to identify a universally optimal kernel, our results instead reveal the fundamentally task-dependent nature of kernel performance. These findings demonstrate the remarkable sensitivity of cosmological parameter estimation to subtle variations in kernel formulation, emphasizing the critical need for kernel pre-selection aligned with specific research objectives—be it $\chi^2$ minimization, bias reduction, marginal likelihood maximization, or other relevant criteria. Moreover, kernel selection uncertainty should be incorporated into cosmological error budgets through frameworks that marginalize over kernel choice, thereby accounting for methodological uncertainties.  Prioritizing kernels based on physical expectations—such as enforcing smooth expansion histories for standard cosmology or allowing greater flexibility to detect new physics—can further improve robustness and interpretability.

While our investigation focused on basic two-parameter kernels, we acknowledge that more sophisticated architectures -- including composite kernels or higher-parameter formulations—may prove advantageous for particular applications. Future work should explore these alternatives while incorporating complementary selection methodologies such as Bayesian model averaging \citep{duvenaud2013structure}, cross-validation techniques \citep{sundararajan1999predictive}, and automated kernel composition approaches \citep{kim2018scaling}. Notably, when complete covariance matrices become available, our modeling framework stands to benefit \rep{substantially}{significantly} from their incorporation, potentially yielding improved statistical precision and more robust parameter constraints.

The $\log(z+1)$ transformation proves particularly promising for cosmological reconstruction, especially given the superior measurement precision and reduced systematic uncertainties characteristic of low-redshift data compared to high-redshift observations. This approach may enhance the theoretical consistency between reconstructed results and established cosmological frameworks—including \del{$\Lambda$CDM,} dynamical dark energy models, and modified gravity theories—potentially delivering superior reconstruction fidelity and tighter cosmological constraints.

The expanding application of GP in cosmological studies necessitates rigorous kernel function optimization, as this choice fundamentally impacts the fidelity of parameter reconstruction and subsequent physical interpretations. Looking forward, as precision cosmology and multi-messenger astronomy enter a new era of increasingly rich and complex data, Gaussian process methods—equipped with robust kernel selection and uncertainty quantification frameworks—will play a pivotal role in extracting unbiased, high-fidelity cosmological information and probing fundamental physics.

\acknowledgments
\add{We are grateful for the referee’s  insightful and useful comments, which greatly helped us improve our manuscript. We thank Wei Hong, Jing Niu, Yu-chen Wang and Hao Zhang for useful discussions.} This work was supported by National Key R\&D Program of China, No.2024YFA1611804, the China Manned Space Program with grant No.CMS-CSST-2025-A01, National Natural Science Foundation of China (NSFC) under the Youth Program (No. 12403004), the Postdoctoral Fellowship Program (Grade C) of China Postdoctoral Science Foundation(GZC20241563) and the national Key Program for Science and Technology Research Development (No. 2023YFB3002500). 

% \paragraph{Note added.} We thank the AI tool (DeepSeek Chat) for its assistance in polishing the language of this manuscript.


% \begin{thebibliography}{99}

% \bibitem{a}
% Author, \emph{Title}, \emph{J. Abbrev.} {\bf vol} (year) pg.

% \bibitem{b}
% Author, \emph{Title},
% arxiv:1234.5678.

% \bibitem{c}
% Author, \emph{Title},
% Publisher (year).

% % Please avoid comments such as "For a review'', "For some examples",
% % "and references therein" or move them in the text. In general,
% % please leave only references in the bibliography and move all
% % accessory text in footnotes.

% % Also, please have only one work for each \bibitem.

% \end{thebibliography}

\bibliographystyle{JHEP}
\bibliography{main}

\end{document}